\documentclass[10pt,twocolumn,english,preprint]{elsarticle}

\usepackage{graphicx}
\usepackage[english]{babel}
\usepackage{amsmath}
\usepackage{amssymb}
\usepackage{amsfonts}
\usepackage{longtable}
\usepackage{dcolumn}
\usepackage{bm}
\usepackage{bbm}
\usepackage{color}
\usepackage{array}
\usepackage{colortbl}

\usepackage{bbold}
\usepackage{ctable}

\usepackage{natbib}

\journal{Journal of Magnetic Resonance}

\begin{document}

\begin{frontmatter}

\title{Modulated pulses compensating classical noise}

\author[do]{Christopher Stihl}
\ead{christopher.stihl@tu-dortmund.de}
\address[do]{Lehrstuhl f\"{u}r Theoretische Physik I, TU Dortmund, Otto-Hahn Stra\ss{}e 4, 44221 Dortmund, Germany}

\author[do]{Benedikt Fauseweh}
\ead{benedikt.fauseweh@tu-dortmund.de}

\author[fzj]{Stefano Pasini,\corref{cor1}}
\ead{s.pasini@fz-juelich.de}
\address[fzj]{Forschungszentrum J\"ulich, 52425 J\"ulich, Germany} 

\author[do]{G\"otz S. Uhrig,\corref{cor1}}
\ead{goetz.uhrig@tu-dortmund.de}

\cortext[cor1]{Corresponding authors}

\begin{abstract}
We consider pulses of finite duration for coherent control in the presence of classical 
noise. We derive the corrections to ideal, instantaneous pulses
for the case of general decoherence (spin-spin relaxation and spin-lattice relaxation)
up to and including the third order in the duration 
$\tau_{\text p}$ of the pulses. For pure dephasing (spin-spin relaxation only), we design 
$\pi$ and $\pi/2$ pulses with amplitude and/or frequency 
modulation which resemble the ideal ones up to and including the second
order in $\tau_{\text p}$. For completely general decoherence including spin-lattice
 relaxation the corrections are computed up to and including the second order in
  $\tau_{\text p}$ as well. 
Frequency modulated pulses are determined which resemble the ideal ones.
They are used to design a low-amplitude replacement for XY8 cycles.
In comparison with pulses designed to compensate quantum noise
less conditions have to be fulfilled. Consequently, we find that the classical pulses
can be weaker and simpler than the corresponding pulses in the quantum case.
\end{abstract}



\maketitle

\end{frontmatter}

\section{Introduction}

In open quantum systems the interaction with an environment is one of the main
reasons for the decay of the coherence and for the loss of
information, both obstacles to an efficient processing of quantum
information or to high precision NMR measurements.  
But by means of a proper time-dependent modulation of the system dynamics,  
it is possible to average out the effects of the environment.
This leads to the prolongation of the coherence time and to the suppression
of the decoherence.

In NMR, a standard way of time-dependent modulation is the application of
very short pulses which flip spins. This approach dates back to 
the spin echo of Erwin Hahn in the fifties \cite{hahn50} and then was
quickly extended to periodically applied pulses by Carr and Purcell
\cite{carr54} and Meiboom and Gill \cite{meibo58}. Since then a large variety
of periodic pulse sequences has been proposed \cite{haebe76,levit01} 
among which we highlight the XY4 cycle \cite{mauds86,gulli90} 
and its derivatives as a means to preserve all three spin directions.

Recently, the approach of stroboscopic pulsing of two-level systems
to suppress their decoherence has been revived under the name of
dynamic decoupling (DD) for
open quantum systems in the context of quantum information processing (QIP)
\cite{viola98,ban98,viola99a}. Again, the idea is to effectively decouple
the two-level system from the environment which induces its decoherence.
Interestingly, quantum information processing requires a slightly differing
performance than NMR. During the information processing it is desired that 
the decoherence is much better suppressed to the level of $10^{-2}$ to $10^{-4}$. 
But deviations at longer times do not matter. In contrast, in NMR a loss of signal of
the order of 50\% still allows one to track the signal, but it is desired to
be able to do this for as long as possible.

Hence, in QIP novel sequences of pulses have been recently developed which
do not rely on periodically repeated cycles, but
vary the time intervals between the pulses. A first suggestion relies
on recursively concatenated sequences (CDD). They require an exponentially growing
number of pulses \cite{khodj07} if longer times have to be reached. 
For the suppression of dephasing (spin-spin relaxation) the Uhrig dynamic decoupling (UDD) 
is much more efficient  because it requires only a linearly growing number of pulses 
\cite{uhrig07,uhrig07err,uhrig08,yang08}. In practice, this pays for 
power spectra with hard cutoff \cite{cywin08,uhrig08,pasin10a}.
Subsequently, UDD has been extended 
to also suppress general decoherence including spin-lattice relaxation due to 
spin flips. This sequence goes under the name of quadratic dynamic decoupling (QDD)
\cite{west10,pasin10b,quiro11,kuo11} and it is further generalized to nested
UDD to deal with more than one two-level system, i.e., with quantum registers
\cite{wang11b}. Experimental verification of these theoretical proposals
are also available for the UDD \cite{bierc09a,du09}.

All the above approaches rely theoretically on ideal, instantaneous pulses
which can not be realized experimentally. There are various routes to overcome
this problem. One is to abandon the use of pulses altogether by resorting
to continuous control fields, see for instance Ref.\ \cite{gordo08}.
We will not follow this route here but stick to sequences of pulses.
We point out, however, that there is a crossover between both approaches if the pulses become
so long that one touches the next one or if a pulse is designed such that 
it acts itself like a cycle of pulses, see below.

For pulses, the most direct solution is to realize very short pulses
so that their duration $\tau_\text{p}$ is so short that the coupling
to the bath does not play a role during the pulse
\begin{equation}
\label{eq:aim1}
U_\text{p} \left(\tau_\text{p}, 0\right) = \hat{P}_{\tau_\text{p}} + {\cal O}(\tau_\text{p})
\end{equation} 
where $U_\text{p} \left(\tau_\text{p}, 0\right)$ is the unitary time evolution
under the combined action of the coupling between system (spin, qubit, or
two-level system) and bath as well 
as the coherent control while $\hat{P}_{\tau_\text{p}}$ stands for the time
evolution under the pulse alone as if the system were isolated.

A more sophisticated solution consists in shaping the pulse such that
the above approximation holds in the form
\begin{equation}
\label{eq:aim2}
U_\text{p} \left(\tau_\text{p}, 0\right) = \hat{P}_{\tau_\text{p}} + {\cal O}(\tau_\text{p}^{m+1}).
\end{equation} 
We will call a pulse which fulfills \eqref{eq:aim2}
an $m$-th order pulse. Thus, a standard unshaped pulse is a 0-th order pulse.
In this work, we will present first and second order pulses.

We emphasize that pulses with the property \eqref{eq:aim2} cannot be used as
simple replacements for an ideal, instantaneous pulse because they
still last for the time $\tau_\text{p}$. A proper replacement would have to fulfill
\begin{equation}
\label{eq:aim3}
U_\text{p} \left(\tau_\text{p}, 0\right) = U\left(\tau_\text{p}, \tau_\text{id}\right)\hat{P}_{\tau_\text{p}}
U\left(\tau_\text{id},0\right) + {\cal O}(\tau_\text{p}^{m+1})
\end{equation} 
where $U\left(\tau_2, \tau_1\right)$ is the unitary time evolution without any external
control and $\tau_\text{id}$ is the instant at which the $\delta$-like control pulse
$\hat{P}_{\tau_\text{p}}$ is inserted. The time evolutions $U$ before and after an
ideal, instantaneous pulse $\hat{P}_{\tau_\text{p}}$ simulate that right before and after an ideal pulse
the system is coupled to its decohering environment.
It could be shown, however, that for $\pi$ pulses
the ansatz \eqref{eq:aim3} can at most be fulfilled
for $m\le 1$ so that the route of finding proper replacements cannot be pursued further
\cite{pasin08a,pasin08b}.

Hence, the ansatz \eqref{eq:aim2} had to be developed further because
it can be realized for arbitrary $m$ in principle \cite{khodj10} and concrete
proposals have recently been made for second order pulses ($m=2$)
for general quantum baths \cite{pasin09a,fause12}. It could be shown
that pulses with this property can be used in sequences of pulses if they are not
used in the same manner as instantaneous pulses would be used, but with adapted sequences
\cite{uhrig10a,pasin11a}. These observations clearly show that the shaping of pulses
is an important ingredient for high-fidelity coherent control.

Of course, shaping of pulses has also been a long-standing issue in NMR. The main aim
was to generate robust pulses which are only weakly susceptible to pulse imperfections.
We can mention only partly the abundant literature on this issue
\cite{tycko83,levit86,cummi00,cummi03,skinn03,kobza04,sengu05,motto06,alway07,pryad08a,pryad08b},
for a book see Ref.\ \cite{levit05}.

In contrast to earlier work, e.g., the one by Skinner and collaborators \cite{skinn03,kobza04}, 
we focus in this article on analytical derivations as far as possible. In contrast to our 
previous work on pulses for systems coupled to quantum baths \cite{pasin09a,fause12}
we will discuss classical baths. The fundamental reason is that
in most experiments the dominating fluctuations destroying coherence are of \emph{classical}
nature, see, e.g., Ref.\ \cite{bierc11b}. Generically the decohering fluctuations 
are induced by a large number of microscopic and macroscopic degrees of freedom.
These degrees of freedom are at least partly at rather high temperatures relative to their 
generic energy scales. For instance, the nuclear spins can mostly be considered
to be in a disorderd state corresponding to infinite temperature. 
Thus the resulting fluctuations are thermal fluctuations and their quantum character
plays only a smaller role.

One may think that the strongly disordered states of the decohering baths pose a problem
to the preservation of coherence in the systems under study. But the opposite is true:
The lack of quantumness of the fluctuations allows us to consider classical fluctuations only.
This implies less restrictive conditions on $m$-th order pulses of the type \eqref{eq:aim2}.
Hence, the necessary pulse shapes are simpler and most importantly the required amplitudes
are lower than in the full quantum case. The main 
goal of the present article is to show this explicitly. 
The concomitant message to experiment is that for given power supply the classical pulses
will be easier to realize. They can be made shorter than their quantum analogues.

In this article, we derive  amplitude and frequency
modulated pulses for pure dephasing due to classical noise, 
i.e., for spin-spin relaxation without spin flips, up to and including second order ($m=2$). 
Particular attention will be paid to the minimization of the amplitude
because this is important for practical use.
For general decoherence due to classical noise, i.e., 
including spin flips due to spin-lattice relaxation, we derive frequency
modulated pulses up to and including second order ($m=2$). 
We stress that such pulses can be used to build continuous versions of the well-known
XY8 cycle \cite{mauds86,gulli90}.

The paper is organized as follows: In section \ref{sec:geneq} the equations
for the decoupling of general decoherence are derived.  In sections
\ref{sec:AM_dephasing} and \ref{sec:FM_pulses} these equations are
shown for the first and second order. They are solved
numerically for the case of pure dephasing, both for
amplitude and for frequency modulated pulses. In section \ref{sec:AM_FM_pulses} we study
decoupling pulses that combine the characteristics of amplitude (AM) and
frequency modulation (FM) in order to illustrate that transient amplitudes, when
the pulse is switched on or off, can be easily dealt with.
In section \ref{sec:gen_decoherence} the case
of general decoherence is solved for by frequency modulated pulses. 
We draw our conclusions in section \ref{sec:conclusions}.

\section{General Equations}
\label{sec:geneq}

\subsection{Ansatz}
\label{subsec:ansatz}

We consider the following Hamiltonian
\begin{equation}
\label{eq:hamiltonian}
	H_\text{tot}\left(t\right) = H\left(t\right) + H_0\left(t\right)
\end{equation} 
which consists of two parts: A system Hamiltonian
\begin{equation}
\label{eq:Hsystem}
  H\left(t\right)=\vec{\eta}\left(t\right)\cdot\vec{\sigma}
  \end{equation} 
and a control Hamiltonian 
\begin{equation}
\label{eq:Hcontrol}
H_0\left(t\right)=\vec{v}\left(t\right)\cdot\vec{\sigma}.
\end{equation}

The system Hamiltonian describes the physics of a single spin coupled to a classical
bath through the time-dependent random function $\vec{\eta}(t)$. 
We describe decoherence by averaging over $\vec{\eta}(t)$. One may
assume $\vec{\eta}(t)$ to represent Gaussian fluctuations, but this is not
essential for the present work. The vector $\vec\sigma$  with components given by
the Pauli matrices $\sigma_x$, $\sigma_y$ and $\sigma_z$ stands for the spin operator.
The spin is subject to general dephasing including spin-lattice relaxation
if all three vector components are present.

In the control Hamiltonian $H_0(t)$, 
$\vec{v}(t)$ is the time-dependent vector-valued amplitude of the pulse operating on 
the spin. At each instant in time, the spin undergoes a rotation about the  
time-dependent axis $\vec{v}(t)$.

In all what follows we assume that the control $|\vec{v}|$ is large enough 
$\pi \approx \vec{v}\tau_\text{p}$ so that 
$\tau_\text{p}$ can be chosen small enough and 
$\pi \gtrapprox \|H_{0}\|\tau_\text{p}$ holds. Then it is well
justified to expand in the decoherence dynamics. Formally, we perform
an expansion in $\tau_\text{p}$ while  $|\vec{v}|\tau_\text{p}$ is kept at order unity.

Following the technique developed in Refs. \cite{pasin09a,fause12} we split
the complete time-evolution operator 
$U_\text{p} \left(\tau_\text{p}, 0\right) = {\cal T} \left\{ \exp[-i
\int\limits_{0}^{\tau_\text{p}} H_{\text{tot}}\left(t\right)\mathrm{dt}] \right\}$ 
of the system during a pulse of duration $\tau_\text{p}$ into two terms: (i) The
evolution of the spin under the effect of only the control field $\hat{P}_{\tau_\text{p}}$ -- as
if it were decoupled from the bath --
and (ii) the correction  $U_\text{c}\left(\tau_\text{p}, 0\right)$
\begin{equation}
\label{eq:ansatz}
U_\text{p} \left(\tau_\text{p}, 0\right)  
:= \hat{P}_{\tau_\text{p}} U_\text{c}\left(\tau_\text{p}, 0\right).
\end{equation} 
The notation $\cal T$ stands for the quantum mechanical time-ordering
operator. The operator $U_\text{c}\left(\tau_\text{p}, 0\right)$ incorporates the
deviations from an ideal rotation resulting from the interaction
between the spin and the bath. 

The rotation operator can be expressed as
\begin{equation}
\label{eq:rot}
\hat{P}_{t}:=\exp\{ -i\vec{\sigma}\cdot\vec{a}(t)\frac{\psi(t)}{2}\},
\end{equation}     
where the unit vector $\vec a$ stands for the axis of total rotation
while $\psi(t)$ is the corresponding angle by which the spin
is rotated from the instant $\tau=0$ till the instant $\tau=t$. 

Both operators
$\hat P_t$ and $U_\text{p} \left(t, 0\right)$ satisfy a Schr\"odinger equation
at every instant $t$, for details see Ref.\ \cite{pasin09a,fause12}. 
For the former operator $i\partial_t\hat P_t=H_0(t)\hat P_t$ 
translates to 
 \begin{align}
2 \vec{v}(t) &=  \psi'(t) \hat{a}(t) +  \hat{a}'(t) \sin \psi(t) 
\nonumber \\
 &- (1- \cos \psi(t))\left[ \hat{a}'(t) \times \hat{a}(t) \right] .
\label{eq:vdes} 
\end{align} 
For the latter, one obtains the formal solution for $t\in [0,\tau_{\text p}]$
\begin{equation}
\label{eq:solSchroedinger}
U_\text{c}(\tau_\text{p},0) =T\left\{e^{-i\int_0^{\tau_\text{p}} \tilde{H}(t)\text{d}t} \right\},
\end{equation} 
where 
\begin{equation}
\label{eq:tildeH_def}
\tilde H(t):=\hat{P}_{t}^{-1}H\left(t\right)\hat{P}_{t}.
\end{equation}

\subsection{Integral Equations}
\label{subsec:int_equations}

One of the standard procedures to deal with time-dependent Hamilton operators as in 
\eqref{eq:solSchroedinger} is the Magnus expansion \cite{magnu54,blane09}.
It consists in replacing the time
dependent Hamiltonian by a sum of average time-independent
Hamiltonian terms 
\begin{equation}
\label{eq:magnus}
U(0,\tau_\text{p}) =\exp\left\{-i\tau_\text{p} \sum_{n}\tilde{H}_n\right\},
\end{equation} 
where $\tilde{H}_n={\cal O}(\tau_\text{p}^{n-1})$.
The explicit form of the first three cumulants of
the expansion is given by
\begin{subequations}
\label{eq:Magnus}
\begin{align}
\label{eq:Magnus1}
\tau_\text{p} \tilde{H}_1&=\int_0^{\tau_\text{p}} \text{d}t\tilde{H}(t)\\
\label{eq:Magnus2}
\tau_\text{p} \tilde{H}_2&=-\frac{i}{2}\int_0^{\tau_\text{p}}
\text{d}t_1 \int_0^{t_1}
\text{d}t_2\left[\tilde{H}(t_1),\tilde{H}(t_2)\right]\\
\tau_\text{p} \tilde{H}_3&=-\frac{1}{6}\int_0^{\tau_\text{p}}
\text{d}t_1 \int_0^{t_1}
\text{d}t_2 \int_0^{t_2}
\text{d}t_3\notag \\ 
\notag
&\left\{\left[\left[\tilde{H}(t_1),\tilde{H}(t_2)\right],\tilde{H}(t_3)\right]\right.
\\
& +\left.\left[\tilde{H}(t_1),\left[\tilde{H}(t_2),\tilde{H}(t_3)\right]\right]\right\}.
\label{eq:Magnus3}
\end{align}
\end{subequations}

The rotated spin operator $\hat{P}_{t}^\dag\vec{\sigma}\hat{P}_{t}$ can 
also be understood as the vector $\vec{\sigma}$ rotated by the angle $\psi$ about the
axis $\vec{a}(t)$. 
By applying this rotation matrix not to $\vec\sigma$, but its inverse to $\vec\eta$
we can reexpress the Hamiltonian as $\tilde{H}\left(t\right) =
\vec{n}_{\vec{\eta}}\left(t\right)\cdot\vec{\sigma}$, where
$ \vec{n}_{\vec{\eta}}\left(t\right) :=
\left[D_{\vec{a}}\left(-\psi(t)\right)\cdot\vec{\eta}\right]$. 
The corresponding $3\times 3$ rotation matrix $D_{\vec{a}}\left(\psi\right)$
is given in \ref{app:rotmatrix}.
The rotated vector $\vec \eta$ has the following explicit form \cite{pasin09a,fause12}
\begin{align}
\label{eq:eta}
\notag
	\vec{n}_{\vec{\eta}}\left(t\right)  &=
        \cos\left(\psi\right) \vec{\eta} -
        \sin\left(\psi\right)\left[\vec{a}\times\vec{\eta}\right]\\ &
        +
        \left[1-\cos\left(\psi\right)\right]\left(\vec{\eta}\cdot\vec{a}\right)
        \vec{a} 
\end{align}
where we did not denote the time dependence of $\psi, \eta$, and $\vec{a}$ 
for the sake of clarity.

Now we are able to derive an expression for the
terms of the Magnus expansion
\begin{subequations}
\begin{align}
\label{eq:Magnus1_n}
	\tau_\text{p}\tilde H_1 = 
	\int\limits_{0}^{\tau_\text{p}}\mathrm{dt}\ \vec{n}_{\vec{\eta}}(t)\cdot\vec{\sigma},
\end{align}
\begin{align}
\label{eq:Magnus2_n}
	\tau_\text{p}\tilde H_2 =
\int\limits_{0}^{\tau_\text{p}}\mathrm{dt_{1}}\int\limits_{0}^{t_{1}}
\mathrm{dt_{2}}\  \vec{\sigma}\left(\vec{n}_{\vec{\eta}}\left(t_{1}\right)\times\vec{n}_{\vec{\eta}}\left(t_{2}\right)\right).
\end{align} 
\begin{align} \notag
	\tau_\text{p}\tilde H_3
	= &
    \frac{2}{3}\vec{\sigma}\int\limits_{0}^{\tau_\text{p}}\mathrm{dt_{1}}\int\limits_{0}^{t_{1}} 
    \mathrm{dt_{2}}\int\limits_{0}^{t_{2}} \mathrm{dt_{3}}
  \\ \notag& \left[2\vec{n}_{\vec{\eta}}\left(t_{2}\right)\left(\vec{n}_{\vec{\eta}}\left(t_{1}\right)\vec{n}_{\vec{\eta}}\left(t_{3}\right)\right)\right.
  \\  & -\vec{n}_{\vec{\eta}}\left(t_{3}\right)\left(\vec{n}_{\vec{\eta}}\left(t_{1}\right)\vec{n}_{\vec{\eta}}\left(t_{2}\right)\right)
\notag \\ &
	 -\left. \vec{n}_{\vec{\eta}}\left(t_{1}\right)\left(\vec{n}_{\vec{\eta}}\left(t_{3}\right)\vec{n}_{\vec{\eta}}\left(t_{2}\right)\right) \right].
	 \label{eq:Magnus3_n}
\end{align}
\end{subequations}
Clearly, if 
\begin{align}
\label{eq:conditionMagnus}
\tilde H_1 = \tilde H_2 =\tilde H_3 =0,
\end{align} 
 holds, the corresponding pulse fulfills \eqref{eq:aim2}
 for $m=3$. If only $\tilde H_1$ and $\tilde H_2$ vanish then
 \eqref{eq:aim2} holds  for $m=2$. 

But we do not know the time-dependent $\eta(t)$. As usual in statistical
physics we average over all possible realizations $\{\eta(t)\}$ for
which we need the correlations of $\{\eta(t)\}$. But it is in general
not sufficient to average $U_\text{c}$ over $\{\eta(t)\}$.
One has to average the changes in the density matrix $\hat\rho$ of the quantum spin.
Thus we consider $U_\text{p}^\dag\hat\rho U_\text{p}$ and average this
quantity over $\{\eta(t)\}$
\begin{equation}
U_\text{p}^\dag\hat\rho U_\text{p} \to \langle U_\text{p}^\dag\hat\rho U_\text{p}\rangle_\eta
\end{equation}
as indicated by $\langle \cdot \rangle_\eta$. 
For $S=1/2$, i.e., the two-level quantum system under study, $\hat\rho$
can be expanded in the Pauli matrices and the identity. The latter
is transformed trivially so that it is sufficient to consider
\begin{equation}
\langle U_\text{p}^\dag\vec{\sigma} U_\text{p}\rangle_\eta
= \langle U_\text{c}^\dag\vec{\sigma}' U_\text{c}\rangle_\eta
\end{equation}
where $\vec{\sigma}'=\hat P_{\tau_\text{p}}^\dag \vec{\sigma} \hat P_{\tau_\text{p}}$.
To order $m$ we require that 
\begin{equation}
\label{eq:aim}
 \langle U_\text{c}^\dag \vec{\sigma}' U_\text{c}\rangle_\eta = \vec{\sigma}'
 +{\cal O}(\tau_\text{p}^{m+1}).
\end{equation}
The vector $\vec{\sigma}$ parametrizes all deviations of the density matrix
from the identity and so does $\vec{\sigma}'$. Hence we need not distinguish
$\vec{\sigma}$ and $\vec{\sigma}'$.

In an expansion up to second order in $\tau_{\text p}$ we verified that 
our objective \eqref{eq:aim} is indeed fulfilled if 
$\langle U_\text{c}\rangle_\eta$ is the identity up to terms of the order 
${\cal O}(\tau_\text{p}^{m+1})$. Thus in the sequel, we will consider the average
of the unitary $\langle U_\text{c}\rangle_\eta$. In third order, however,
one would have to consider \eqref{eq:aim} rather than the average of
$U_\text{c}$ alone.
 Next, we will address the solutions for the conditions 
 $\langle \tilde H_i\rangle_\eta=0$.
 To do so we will specify these conditions further.

\section{Amplitude-Modulated Pulses for Pure Dephasing}
\label{sec:AM_dephasing}

Here we consider pure dephasing which means that we exclude spin flips.
Thus we restrict the Hamiltonian of the system to 
\begin{equation}
\label{eq:Hpure_dephasing}
H(t) = \eta(t)\sigma_z.
\end{equation}
In this case  rotations around the $y$ axis in spin space ($\vec a =(0,1,0)$)
are sufficient to decouple the spin from the bath.
Thus the control Hamiltonian may assume the following simple form
\begin{equation}
\label{eq:H0pure_dephasing}
H_0(t) = v(t)\ \sigma_y
\end{equation} 
so that we only have to consider the amplitude modulation of $v(t)$.
Despite its simplicity this approximation is representative for a large class of the
decohering systems in which $T_1$ is much larger than $T_2$. This can be reached
by large magnetic fields which split the two energy levels strongly.
Then all terms different from $\sigma_z$ are averaged out in the rotating frame
approximation.

For $H_0$ as in \eqref{eq:H0pure_dephasing}
the rotation operator $\hat P_t = \cos\left[\psi(t)/2\right]-i\sigma_y
\sin\left[\psi(t)/2\right]$ transforms the system Hamiltonian
(\ref{eq:Hpure_dephasing}) to $\tilde H(t)$ in \eqref{eq:tildeH_def} yielding
\begin{align}
\label{eq:Htilde_pure_dephasing}
\tilde H(t) = \eta(t) \left[\cos(\psi(t))\sigma_z-\sin(\psi(t))\sigma_x\right].
\end{align} 
From Eqs.\ (\ref{eq:eta}), we obtain
\begin{align}
\label{eq:n_pure_dephasing}
\vec n_\eta(t) = \eta(t)\left[ -\sin(\psi(t)), 0,\cos(\psi(t))\right].
\end{align}

Due to the classical nature of the bath only a few of its average values
are needed. In the leading order only the mean value
\begin{align}
\label{eq:mean_eta}
\bar\eta:=\langle \eta(t)\rangle_\eta
\end{align}  
enters. Since we assume the noise to be uniform in time, i.e.,
it does not depend on time, the mean value is a constant.
We find that the vanishing of 
$\tau_\text{p}\tilde H_1 = \mu_{1,1}\sigma_x+\mu_{1,2}\sigma_z$
requires the vanishing of $\mu_{1,i}$ with $i\in\{1,2\}$ with
\begin{subequations}
\label{eq:first_order}
\begin{align}
\label{eq:mu11}
\mu_{1,1} &:= \bar \eta \int_0^{\tau_{\text p}} \text{d}t\
\sin(\psi(t)),\\
\label{eq:mu12}
\mu_{1,2}& := \bar \eta \int_0^{\tau_{\text p}} \text{d}t\ \cos(\psi(t)),
\end{align}
\end{subequations} 
unless $\bar\eta$ is zero accidentally.
As expected the first-order corrections  are of order $\tau_\text{p}$.  

It is straightforward to find solutions of \eqref{eq:first_order}
for piecewise constant amplitudes or continously varying amplitudes.
Since the first order equations \eqref{eq:first_order} are identical to the ones
for the quantum case, we studied and solved them before \cite{pasin09a}.
We refrain from presenting such solutions here again. We stress 
that solutions of  the first order equations \eqref{eq:first_order} 
had been found before in the context of searching for pulses which
are robust against frequency offsets \cite{tycko83,cummi00,cummi03}.

For the second order in $\tau_\text{p}$ we need some information from
the autocorrelation function  
$g(\Delta t):=\langle \eta (t)\eta(t+\Delta t)\rangle_\eta$. 
For short delays $\Delta t \to 0$ we can expand $g(\Delta t)$ in the form
\begin{align}
\label{eq:TaylorAutocorr}
g(\Delta t) = (\bar \eta^2+ s^2) + g_1|\Delta t| + {\cal O}(\Delta t)^2,
\end{align}
where $s$ is the usual variance $s^2=\langle(\eta-\bar\eta)^2\rangle_\eta$.
We do not exclude the appearance of the term proportional to $|\Delta t|$
although it is at odds with analyticity. Often, a very fast microscopic process 
makes $\eta(t)$ behave as in an Ornstein-Uhlenbeck process which is characterized by
a cusp in $g(\Delta t)$ and a Lorentzian power spectrum after Fourier transform, see for
instance Ref.\ \cite{lange10}.
But we will see below that for the purposes of the present article even $g_1$
does not matter for second order pulses. It occurs only in
third order pulses.

Explicit calculation yields the second order correction 
\begin{align}
\label{eq:mu2}\notag
\tau_\text{p} \tilde H_2 :=& \sigma_y  (\bar \eta^2+ s^2)\cdot \\
 &\int_0^{\tau_{\text
    p}}{\text d}t_1\int_0^{t_1}{\text d}t_2 \sin\left[\psi(t_1)-\psi(t_2)\right],
\end{align} 
which is quadratic in $\tau_{\text p}$.

\begin{figure}[htb]
\includegraphics[angle=-90,width=\columnwidth,clip]{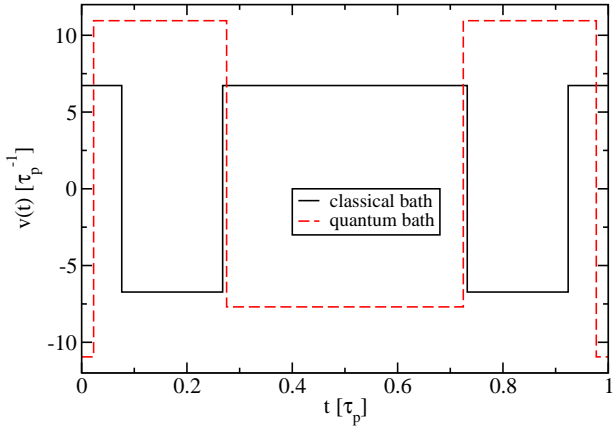}
 \caption{(Color online) Symmetric second order $\pi$ pulse with piecewise constant amplitude suppressing pure dephasing. 
 The symmetric switching instants are $\tau_1 = 0.07623078 
\tau_p$, $\tau_2 = 0.26784319 \tau_p$, $\tau_3 = 1 - \tau_2$ and $\tau_4 
= 1 - \tau_1$, the amplitude is $\pm 6.72572865 \tau_p^{-1}$
 \label{fig:1}}
\end{figure}

\begin{figure}[htb]
 \includegraphics[angle=-90,width=\columnwidth,clip]{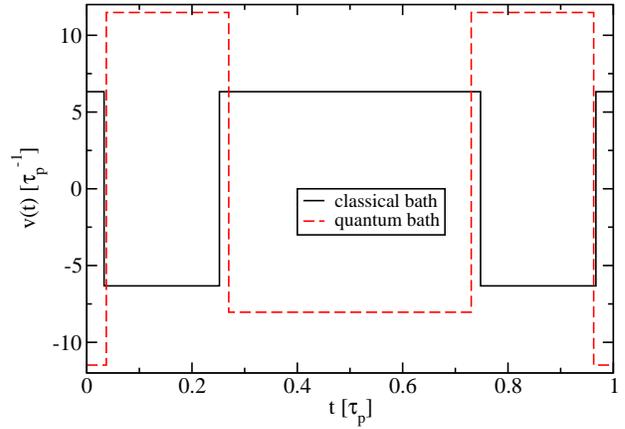}
 \caption{(Color online)  Symmetric second order $\frac{\pi}{2}$ pulse with piecewise constant amplitude  suppressing pure dephasing. 
 The symmetric switching instants are 
$\tau_1 = 0.03312609 \tau_p$, $\tau_2 = 0.25209296 \tau_p$, $\tau_3 = 1 
- \tau_2$ and $\tau_4 = 1 - \tau_1$, the amplitude is $\pm 6.32709469 
\tau_p^{-1}$.
 \label{fig:2}}
\end{figure}

Comparing the integral equations for the first and the second
order with those derived for a quantum bath \cite{pasin09a}
we find that they are identical in first  order. In second order,
they are similar in form but in the quantum case there are three
equations while two of them vanish in the classical case
because the system Hamiltonian (\ref{eq:Hpure_dephasing}) commutes with itself at different times. 
This implies that it is less demanding to find a
numerical solution for the equations \eqref{eq:first_order}
and \eqref{eq:mu2} in the classical case than in the quantum case.
Practically, pulses can be found with a lower amplitude than in
the quantum case. In Figs.\ \ref{fig:1} and \ref{fig:2} examples
of a $\pi$ and a $\pi/2$ second-order pulses with  piecewise constant amplitudes
are shown and compared to the corresponding pulses for a quantum bath \cite{pasin09a}.
Clearly, the maximal amplitudes of the classical pulses are lower.
A technical remark is in order: Although the classical pulses
have to fulfill less conditions we designed them with the
same number of switching instants as the quantum pulses. But
the absolute value of the amplitudes of the classical pulses
is always the same while it varies in the quantum case.
So the amplitude values are the additional variables needed in the quantum case.

\begin{figure} [htb]
 \includegraphics[angle=-90,width=\columnwidth,clip]{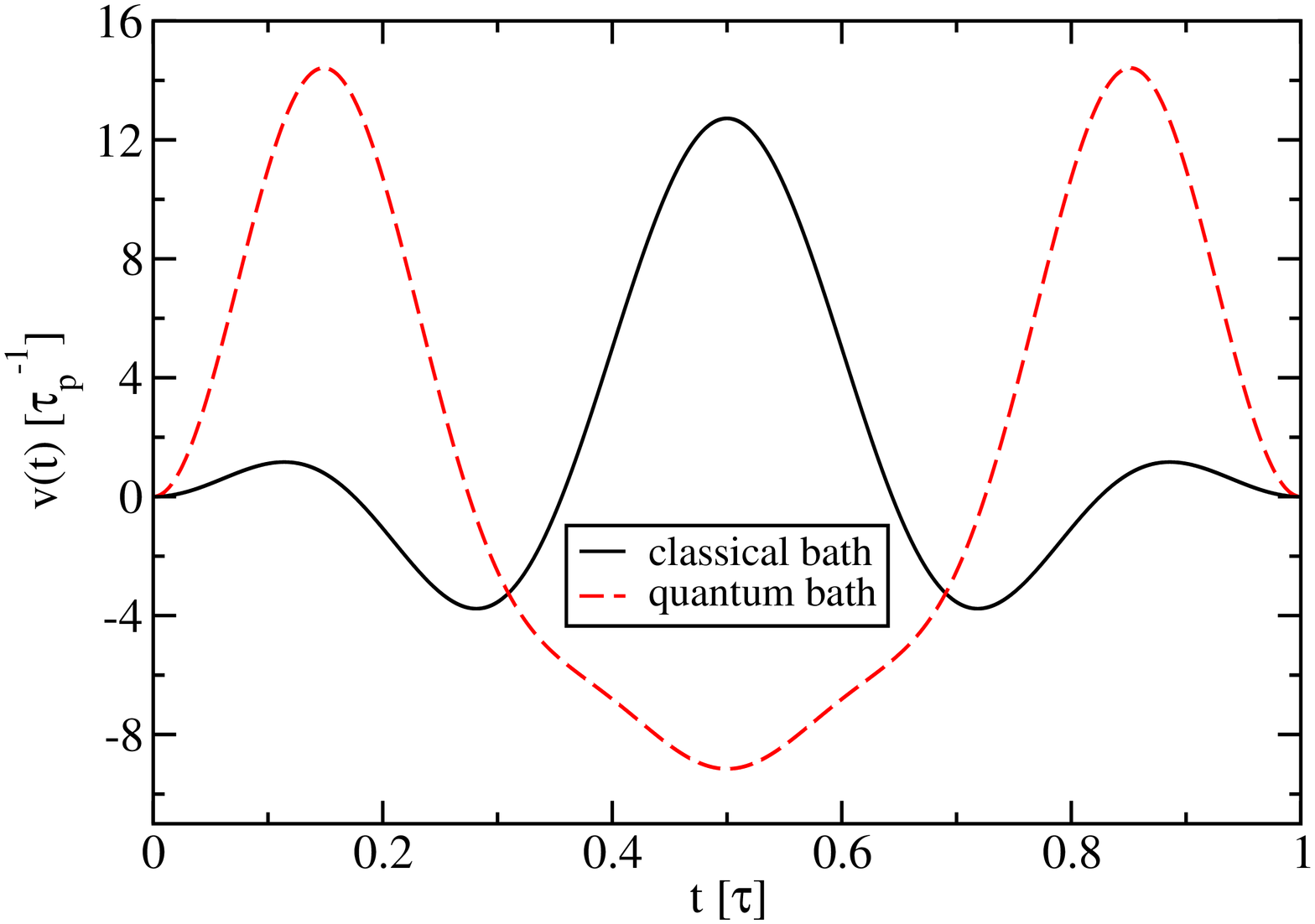}
 \caption{(Color online)  Symmetric $\pi$ pulse with continuosly modulated amplitude correcting up to 
second order suppressing pure dephasing. The function $v(t)$ is parametrized by 
Eq. \eqref{eq:cont_pi_pulse} with $a = -1.92179255$ and $b = 2.86838351$.
 \label{fig:3}}
\end{figure}

\begin{figure}[htb]
 \includegraphics[angle=-90,width=\columnwidth,clip]{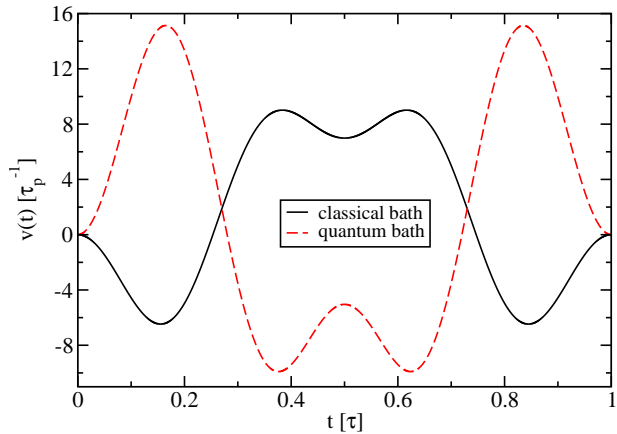}
 \caption{(Color online)  Symmetric $\frac{\pi}{2}$ pulse with
   continuously modulated amplitude 
correcting up to second order  suppressing pure dephasing. The function $v(t)$ is parametrized by 
Eq. \eqref{eq:cont_pi_pulse} with $a = -5.41258549$ and $b = -3.48909926$.
 \label{fig:4}}
\end{figure}

Equations (\ref{eq:first_order},\ref{eq:mu2}) can also be solved for
continuous pulses. For symmetric pulses the function $v(t)$ can be
represented by 
\begin{align}
\label{eq:cont_pi_pulse}\nonumber
v(t)=&
\frac{\theta}{2}+\left(a+\frac{\theta}{2}\right)\cos\left(\frac{2\pi}{\tau_{\text
    p}}t\right)+(b-a) \cos\left(\frac{4\pi}{\tau_{\text
    p}}t\right)\\&-b\cos\left(\frac{6\pi}{\tau_{\text
    p}}t\right)
\end{align} 
where $\theta$ is either $\pi$ or $\pi/2$ and $a$ and $b$ are constants. The pulse
fulfills the requirements $v(0)=v(\tau_{\text p})=0$ and
$v^\prime(0)=v^\prime(\tau_{\text p})=0$, as shown in
Figs.\ (\ref{fig:3}) and (\ref{fig:4}). Also in this case the maximal
amplitude is lower than in the quantum case.

All solutions presented above were obtained numerically. 
The piecewise constant  solutions for amplitude modulation were found using
``fsolve()'' from the ``scipy'' library for Python, which essentially is a wrapper around the ``hybrd'' and ``hybrdj''
algorithms from MINPACK. No further numerical calculations were needed to obtain those solutions because the
integrations were done analytically. This was achieved by partitioning the integration domain into the
intervals between the switching instants.
In this way $\psi(t)$ becomes a linear function within each interval of the partition such that 
the contribution of each interval is analytically available.

To obtain continuous solutions for amplitude modulation, the GNU Scientific Library (GSL) was used for performance
reasons. To find the multidimensional roots of the necessary sets of equations, the ``gsl\_multiroot\_fsolver\_hybrids'' algorithm was used. It is very similar to the ``hybrd'' algorithm in MINPACK except for the internal scaling. 
A proposed solution is accepted if the residue, i.e., the sum over all absolute values $\sum_i\left|f_i\right|$, is less than $10^{-10}$. The integrand values $f_i$ themselves are calculated using ``gsl\_integration\_gaq"
which relies on an adaptive numerical integration using the $61$ point Gauss-Kronrod rules until the estimate of the absolute global error is less than $10^{-12}$. 
To compute multiple integrals, calls of ``gsl\_integration\_gaq" were nested appropriately.

\section{Frequency Modulated Pulses for Pure Dephasing}
\label{sec:FM_pulses}

It is important to
analyze the case of frequency modulated pulses as well because there
may be experimental setups where frequency modulation (FM) can be implemented
much easier or more accurately than amplitude modulation.
There is only an initial and a final jump in the amplitude, the remaining
control evolves smoothly. The initial and final transients will be discussed
in the following section.
Another advantage of frequency modulated pulses is that rotations about
different axes can be realized in a natural way. Thereby, all kinds of deviations from the
initial spin state can be compensated including spin relaxation.
This latter point will be exploited in more detail in section \ref{sec:gen_decoherence}.

We study  pulses acting only in the $xy$-plane of the spin orientation with a 
fixed amplitude $V_0:=\left|\vec{v}\right|$. 
In the rotating frame taking the Larmor frequency $\omega_\mathrm{L}$ into account,
the current axis of rotation is given by
\begin{align}
\label{eq:fm_ansatz}
\vec{v}(t) = \left(\begin{array}{c}V_0 \cos(\Omega(t))
\\
V_0  \sin(\Omega(t))\\0\end{array} \right)
\end{align}
where $\Omega(t)$ is a time dependent phase which is tuned externally. The control Hamiltonian $H_0$ is realized by applying a field perpendicular to the $\sigma_z$-axis 
which is rotating with the Larmor frequency in the laboratory frame
\begin{align}  
\label{eq:fm}
H_0^{\text{lab}}= V_0 \left[\sigma_x \cos (\omega_\mathrm{L} t - \Omega(t))
 - \sigma_y \sin (\omega_\mathrm{L} t - \Omega(t))\right].
\end{align}
The derivative $\partial_t \Omega(t)$ of the time dependent phase 
represents the deviation of the current frequency from the Larmor frequency.
In this sense Eq.\ \eqref{eq:fm} describes a frequency modulated pulse.
In the rotating frame we have
\begin{align} 
\label{eq:FMH0_ansatz}
H_{0}(t) = \vec{v}(t) \cdot \vec{\sigma} .
\end{align}

In order to find $\hat{a}(t)$ and $\psi(t)$ appearing in the  parametrization
\eqref{eq:rot} of the pulse one has to solve the differential equation \eqref{eq:vdes}. 
Because $\hat{a}(t)$ is a unit vector, it can be parametrized  by two angles
$\phi(t)$ and $\theta(t)$
\begin{align}
\label{eq:a_para}
\hat{a}(t) := \left(\begin{array}{c}a_x(t)\\ a_y(t)\\
    a_z(t) \end{array}\right)
= \left(\begin{array}{c}\sin(\theta(t)) \cos(\phi(t))\\\sin(\theta(t)) \sin(\phi(t))\\\cos(\theta(t))\end{array}\right).
\end{align}
Solving Eq. \eqref{eq:vdes} for the time derivatives of $\psi(t)$, $\phi(t)$, and $\theta(t)$
we finally find
\begin{subequations}
\label{eq:des_result}
\begin{align}
\partial_t \psi &= 2 V_0 \sin \theta \left[ \sin \Omega \sin \phi + \cos \Omega \cos \phi \right]    \label{eq:des_result1}\\
\partial_t \phi &= V_0 \frac{\left[ \cos \frac{\psi}{2} \sin (\Omega - \phi) - \sin \frac{\psi}{2} \cos \theta \cos (\Omega - \phi) \right]}{\sin \frac{\psi}{2} \sin \theta} 
\label{eq:des_result2} \\
\partial_t \theta &= V_0 \frac{\left[ \cos \frac{\psi}{2} \cos \theta \cos (\Omega - \phi) + \sin \frac{\psi}{2} \sin (\Omega - \phi) \right]}{\sin \frac{\psi}{2}}.
  \label{eq:des_result3}
\end{align}
\end{subequations}
The seeming singularities for $\theta=m\pi, m\in \mathbb{Z}$  on the right hand sides of 
Eqs.\ \eqref{eq:des_result2} and \eqref{eq:des_result3} have no \emph{physical} reason. 
They  only result from the choice of spherical coordinates where $\phi$ is ill-defined for 
$\theta=m\pi, m\in \mathbb{Z}$.
In contrast, the global axis of rotation $\hat a$ is ill-defined if
$\psi$ is a multiple of $2\pi$ because then the unitary of the total pulse
is plus or minus the identity so that $\hat a$ could take any direction.
This is reflected in the singuarities of $\psi=2m\pi, m\in \mathbb{Z}$.

At $t=0$ the current axis of rotation $\vec{v}$ and the global one $\vec{a}$
coincide. The former lies by construction in the $xy$-plane implying the inital conditions
\begin{subequations}
\begin{align}
\underset{t \to 0}{\lim} \theta(t) &= \frac{\pi}{2},\\
\underset{t \to 0}{\lim} \psi(t) &= 0,\\
\underset{t \to 0}{\lim} \phi(t) &= \Omega(0) .
\end{align}
\end{subequations}
Note that the last two equations represent our arbitrary choice
which direction is the $x$-direction for the spins and where the phase
$\Omega$ starts. In the numerical solutions below we will use $\Omega(0)=0$.
Inspecting the limit $t\to 0$ one additionally finds
\begin{subequations}
\begin{align}
\partial_t \phi \mid_{t = 0} &= \left. \frac{\partial_t \Omega(t)}{2} \right|_{t = 0}   \\
\partial_t \theta \mid_{t = 0} &= 0  .  
\end{align}
\end{subequations}
The derivative $\partial_t\psi$ follows trivially from Eq.\ \eqref{eq:des_result1}.

We aim at solutions of Eqs. \eqref{eq:Magnus1_n} and \eqref{eq:Magnus2_n}
 both for $\pi$ and $\pi/2$ rotations and for pure dephasing
 $\vec{\eta}\left(t\right) = \left( 0, 0, \eta\left(t\right) \right)$.  
The corresponding vector $\vec n_{\vec \eta}(t)$ read
\begin{align}
\label{eq:deph}\nonumber
& \vec n_{\vec \eta}(t)  \equiv\eta\left(t\right)
\begin{pmatrix}
 n_{x}\left(t\right)\\
 n_{y}\left(t\right)\\
 n_{z}\left(t\right)
\end{pmatrix} \\
&=\eta\left(t\right)
\begin{pmatrix}
 -a_y\sin\left(\psi\right) + \left(1-\cos\left(\psi\right)\right)a_xa_z\\
a_x\sin\left(\psi\right) + \left(1-\cos\left(\psi\right)\right)a_ya_z\\
\cos\left(\psi\right) + \left(1-\cos\left(\psi\right)\right)a_z^2
\end{pmatrix} .
\end{align}  
For the sake of simplicity we have omitted the explicit time dependence of $a$ and $\psi$.

\begin{figure}[htb]
 \includegraphics[angle=0,width=\columnwidth,clip]{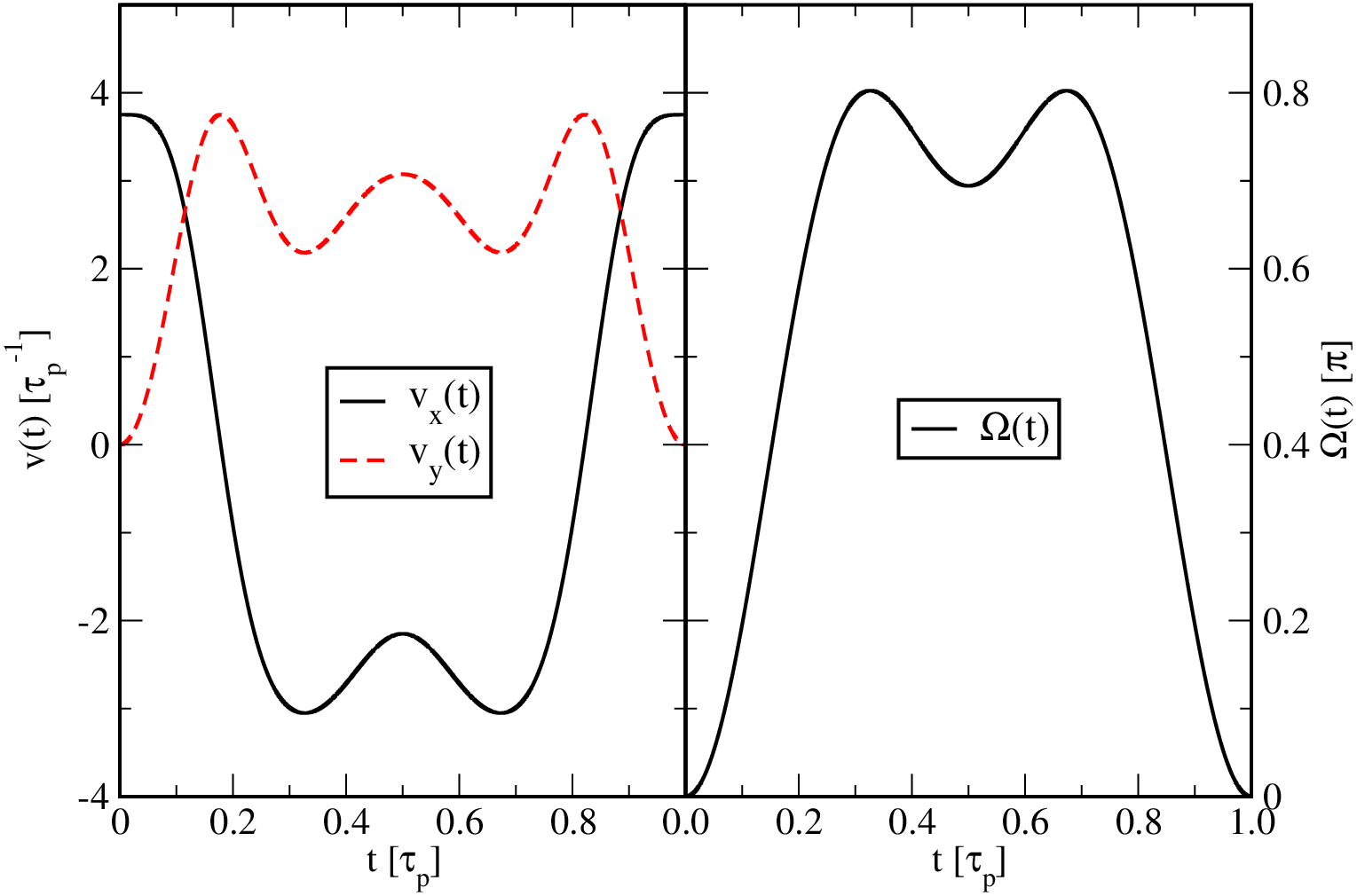}
  \caption{(Color online) First order $\pi$ pulse  suppressing pure dephasing by
   frequency modulation. Parameters given	in Tab.\ \ref{tab:2-FM-DEKO}.}
  \label{fig:1-PI-FM-DEKO}
\end{figure}

\begin{figure}[htb]
 \includegraphics[angle=0,width=\columnwidth,clip]{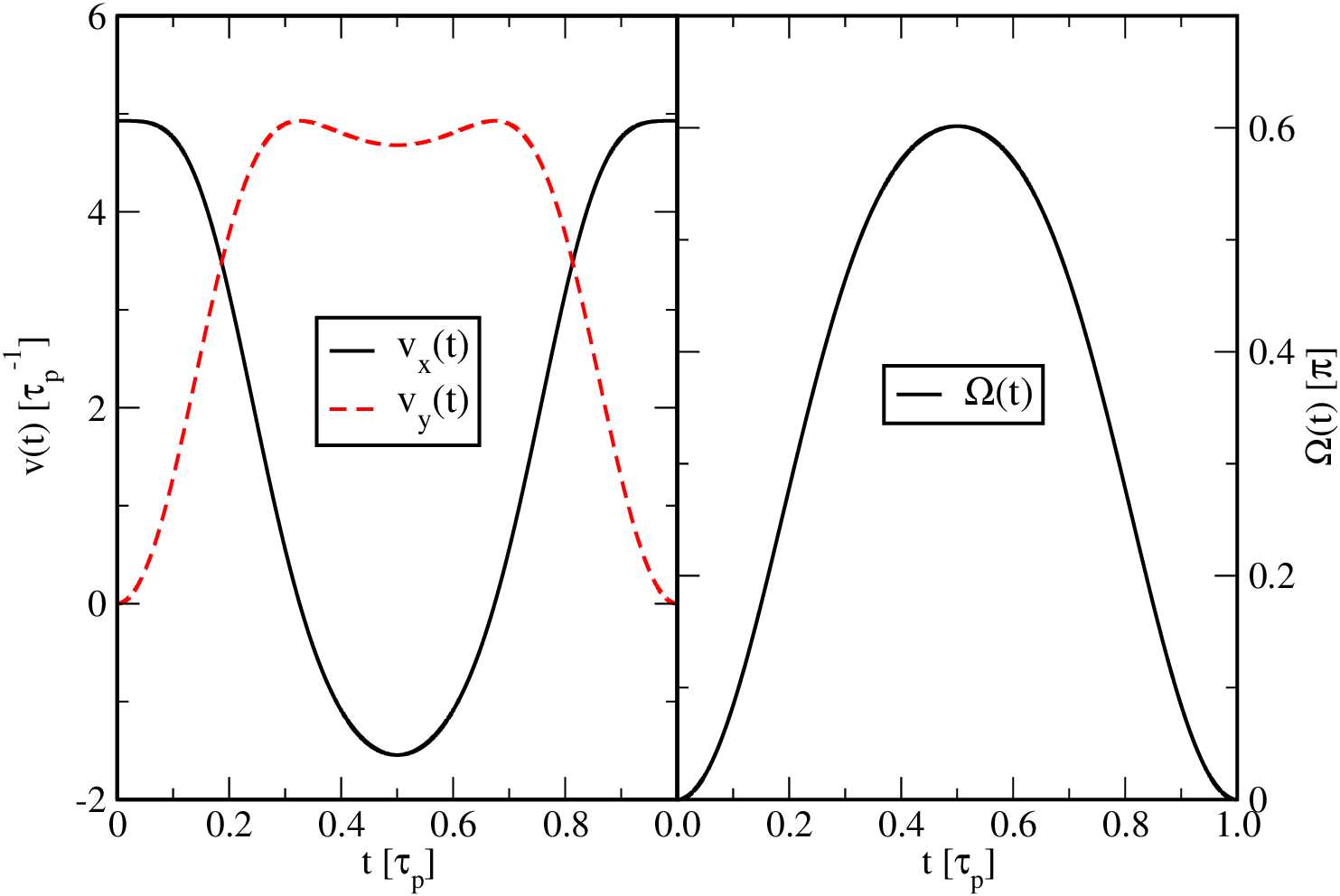}
  \caption{(Color online) First order $\pi/2$ pulse suppressing pure dephasing by
   frequency modulation. Parameters given	in Tab.\ \ref{tab:2-FM-DEKO}.}
  \label{fig:1-PI2-FM-DEKO}
\end{figure}

\begin{table}[htb]
 \centering
 \begin{tabular}{clcl}
\multicolumn{4}{c}{1st order FM $\pi$- and $\frac{\pi}{2}$ pulses}  \\
\multicolumn{2}{c}{$\pi$-pulse} &\multicolumn{2}{c}{$\frac{\pi}{2}$-pulse} \\
\hline								    
$V_0$	&	$3.751157$					&	$V_0$	&	$4.928277$\\
$b_1$	&	$0$	&	$b_1$	&	$0$\\
$b_2$	&	$-1.090479$					&	$b_2$	&	$-0.944852$\\
$b_3$	&	$0$	&	$b_3$	&	$0$\\
$b_4$	&	$-0.588913$					&	$b_4$	&	$-0.122088$\\
\end{tabular}
\caption{Parameters for the first order frequency modulated (FM) pulses suppressing pure dephasing by satisfying Eqs.\ \eqref{eq:1stFM} for compensating a classical bath. 
The coefficients can be compared to those derived in Ref.\ \cite{fause12} for a quantum bath. The dimensionless coefficients $b_n$ refer to the ansatz \eqref{eq:omega_fm}. The amplitudes $V_0$ are given in units of $1/\tau_\mathrm{p}$. The odd coefficients are numerically $0$ within $10^{-13}$.}
\label{tab:2-FM-DEKO} 
\end{table}

For the phase $\Omega(t)$ we use the Fourier series ansatz 
\begin{align}
 \label{eq:omega_fm} 
\Omega(t) &= {\underset{n}{\sum}} b_{2n-1} \sin \big(  \frac{2\pi n
    t}{\tau_{\text{ p}}} \big) 
+ b_{2n} \big[\cos \big( \frac{2\pi n
    t}{\tau_{\text{ p}}} \big)- 1 \big].
\end{align}
The value $\theta(\tau_\mathrm{p})$ is fixed by the fact, that the final axis of rotation has to be 
perpendicular to $\sigma_z$ to rotate the spin by the full angle $\psi(\tau_\mathrm{p})$. Thus one has
\begin{align} 
\label{eq:theta_taup}
\theta(\tau_\mathrm{p}) = \frac{\pi}{2}.
\end{align}
In first order, explicit computation yields 
\begin{equation}
\label{eq:tildeH1}
\tau_\text{p}\tilde H_1=\mu_{1,1}\sigma_x+ \mu_{1,2}\sigma_y+ \mu_{1,3}\sigma_z
\end{equation}
with
\begin{subequations}
\label{eq:eta1_fm_deph}
\begin{align}
	\mu_{1,1} &= \overline{\eta}_z\int\limits_{0}^{\tau_\text{p}}n_{xz}\left(t\right)\mathrm{dt}\\
	\mu_{1,2} &= \overline{\eta}_z\int\limits_{0}^{\tau_\text{p}}n_{yz}\left(t\right)\mathrm{dt}\\
	\mu_{1,3} &= \overline{\eta}_z\int\limits_{0}^{\tau_\text{p}}n_{zz}\left(t\right)\mathrm{dt},
\end{align}
\end{subequations}
where the functions $n_{\alpha\beta}(t)$ with
$\alpha, \beta\in\{x,y,z\}$ are given by the matrix elements
of the rotation matrix $D_{\hat a}(-\psi)$ in \eqref{eq:matrix} in \ref{app:rotmatrix}.
Thus, for first order pulses one has to achieve
\begin{equation}
\label{eq:1stFM}
0=\mu_{1,i}
\end{equation}
for $i\in\{1,2,3\}$. 
Typical solutions for $\pi$ and $\pi/2$ pulses are shown in Figs.\ \ref{fig:1-PI-FM-DEKO}
and \ref{fig:1-PI2-FM-DEKO}. The parameters are given in Tab.\ \ref{tab:2-FM-DEKO}.

In  second order we similarly obtain
\begin{equation}
\tau_\text{p}\tilde H_2=\mu_{2,1}\sigma_x+ \mu_{2,2}\sigma_y+ \mu_{2,3}\sigma_z
\end{equation}
with
\begin{subequations}
\label{eq:eta2_fm_deph}
\begin{align}
\mu_{2,1} = & \int\limits_{0}^{\tau_\text{p}}\int\limits_{0}^{t_{1}} (n_{yz,1}n_{zz,2} - n_{zz,1}n_{yz,2}) \mathrm{dt_{2}}\mathrm{dt_{1}} \\
\mu_{2,2} = & \int\limits_{0}^{\tau_\text{p}}\int\limits_{0}^{t_{1}} (n_{zz,1}n_{xz,2} - n_{xz,1}n_{zz,2}) \mathrm{dt_{2}}\mathrm{dt_{1}} \\
\mu_{2,3} = & \int\limits_{0}^{\tau_\text{p}}\int\limits_{0}^{t_{1}} (n_{xz,1}n_{yz,2} - n_{xz,1}n_{xz,2}) \mathrm{dt_{2}}\mathrm{dt_{1}} ,
\end{align}
\end{subequations}
using the shorthand $n_{\alpha\beta,k}:=n_{\alpha\beta}\left(t_k\right)$ where 
$k\in\{1,2\}$. Thus a second order FM pulse has to fulfill \eqref{eq:1stFM} and
\begin{equation}
\label{eq:2ndFM}
0=\mu_{2,i}.
\end{equation}

\begin{figure}[htb]
 \includegraphics[width=\columnwidth,clip]{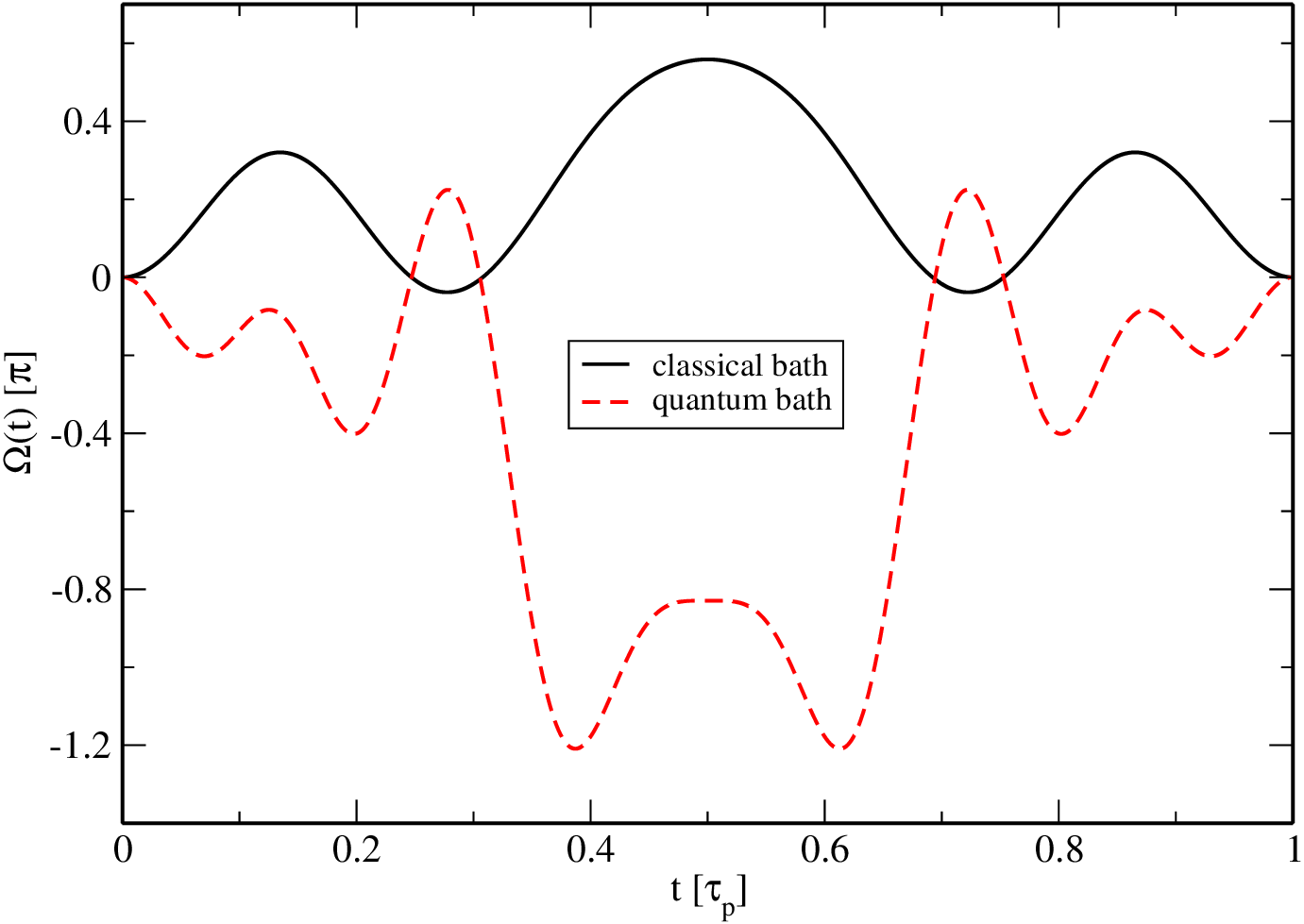}
  \caption{(Color online)  Minimized second order FM $\pi$ pulse suppressing pure
   dephasing with $\Omega(t)$ as in Eq.\ (\ref{eq:omega_fm}). 
   The coefficients for this pulse are given in Tab.\
    \ref{tab:fm_2order_pi}.
  \label{fig5}}
\end{figure}

\begin{figure}[htb]
 \includegraphics[width=\columnwidth,clip]{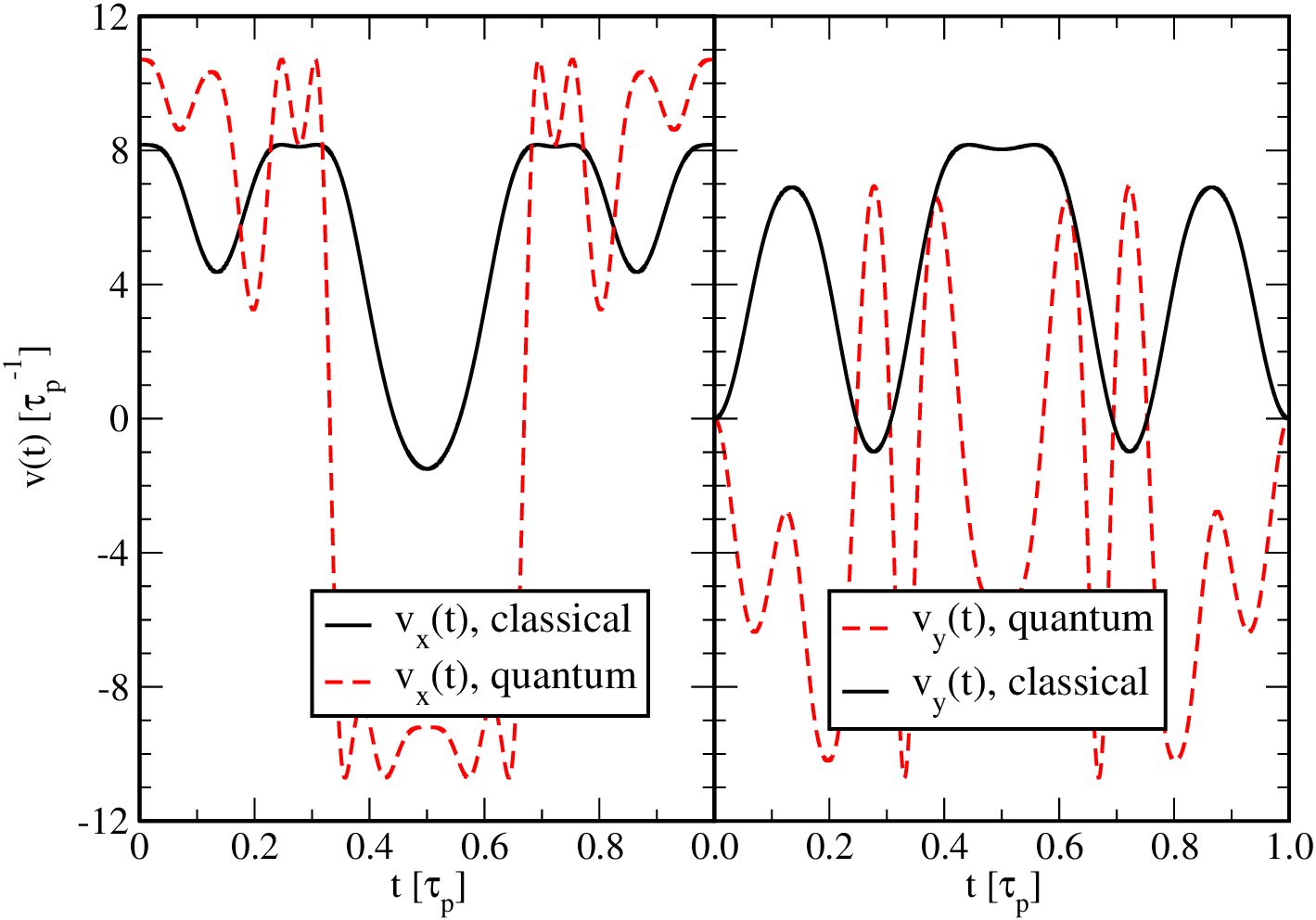}
  \caption{(Color online) Amplitudes for the same pulse as in Fig.\ \ref{fig5}.
  \label{fig6}}
\end{figure}

\begin{table}
 \centering
 \begin{tabular}{clcl}
 \multicolumn{4}{c}{Minimized 2nd order $\pi$ FM pulses}  \\
 \multicolumn{2}{c}{classical} &\multicolumn{2}{c}{quantum}\\
\hline
$V_0$ & $8.129097 $ & $V_0$ & $10.707115$\\
$b_1$ & $0$ & $b_1$ & $0$\\
$b_2$ & $-0.381075$ & $b_2$ & $1.392956$\\
$b_3$ & $0$ & $b_3$ & $0$\\
$b_4$ & $0.450018$ & $b_4$ & $-0.705159$\\
$b_5$ & $0$ & $b_5$ & $0$\\
$b_6$ & $-0.496673$ & $b_6$ & $0.133042$\\
$b_7$ & $0$ & $b_7$ & $0$\\
$b_8$ & $-0.241963$ & $b_8$ & $0.690594$\\
 & & $b_9$ & $0$\\
 & & $b_{10}$ & $-0.695501$\\
 & & $b_{14}$ & $0.472195$\\ 
\end{tabular}
\caption{Parameters of the minimized FM $\pi$ pulses suppressing pure dephasing  by satisfying the first and second order
  Eqs.\ \eqref{eq:1stFM} and \eqref{eq:2ndFM} for a classical or the corresponding
  conditions for a quantum dephasing bath in Ref.\ \cite{fause12}. 
  The dimensionless coefficients $b_n$ refer to Eq.\
  \eqref{eq:omega_fm}. The amplitudes $V_0$ are given in units of
  $1/\tau_\mathrm{p}$. The odd coefficients are numerically $0$ within $10^{-10}$.
\label{tab:fm_2order_pi} }
\end{table}

\begin{table}[htb]
 \centering
 \begin{tabular}{clcl}
\multicolumn{4}{c}{Minimized 2nd order $\pi/2$ FM pulses}  \\
\multicolumn{2}{c}{classic} &\multicolumn{2}{c}{quantum} \\
\hline								    
$V_0$  &  7.405785	&  $V_0$  & 8.435414\\
$b_1$  &  1.524556	&  $b_1$  & -1.820216\\
$b_2$  &  -0.349899	&  $b_2$  & -0.351972\\
$b_3$  &  0.325909	&  $b_3$  & 0.030436\\
$b_4$  &  0.411212	&  $b_4$  & 0.521648\\
$b_5$  &  0.690512	&  $b_5$  & -0.555341\\
$b_6$  &  -0.510771	&  $b_6$  & -0.387557\\
$b_7$  &  0.347745	&  $b_7$  & 0.451462\\
$b_{11}$  &   0.019634	& $b_8$   & -0.193733\\
		&		&	$b_9$		&	-0.161450\\
		&		&	$b_{10}$		&	-0.282067\\
		&		&	$b_{14}$		&	0.047116\\
\end{tabular}
\caption{Parameters of the minimized FM $\pi/2$ pulses suppressing pure dephasing by satisfying all first and second order
Eqs.\ \eqref{eq:1stFM} and \eqref{eq:2ndFM} for a classical bath or the corresponding
  conditions for a quantum bath in Ref.\ \cite{fause12}. 
  The dimensionless coefficients $b_n$ refer to Eq.\
  \eqref{eq:omega_fm}. The amplitudes $V_0$ are given in units of
  $1/\tau_\mathrm{p}$.}
\label{tab:fm_2order_pi2} 
\end{table}

\begin{figure}[htb]
 \includegraphics[angle=-90,width=\columnwidth,clip]{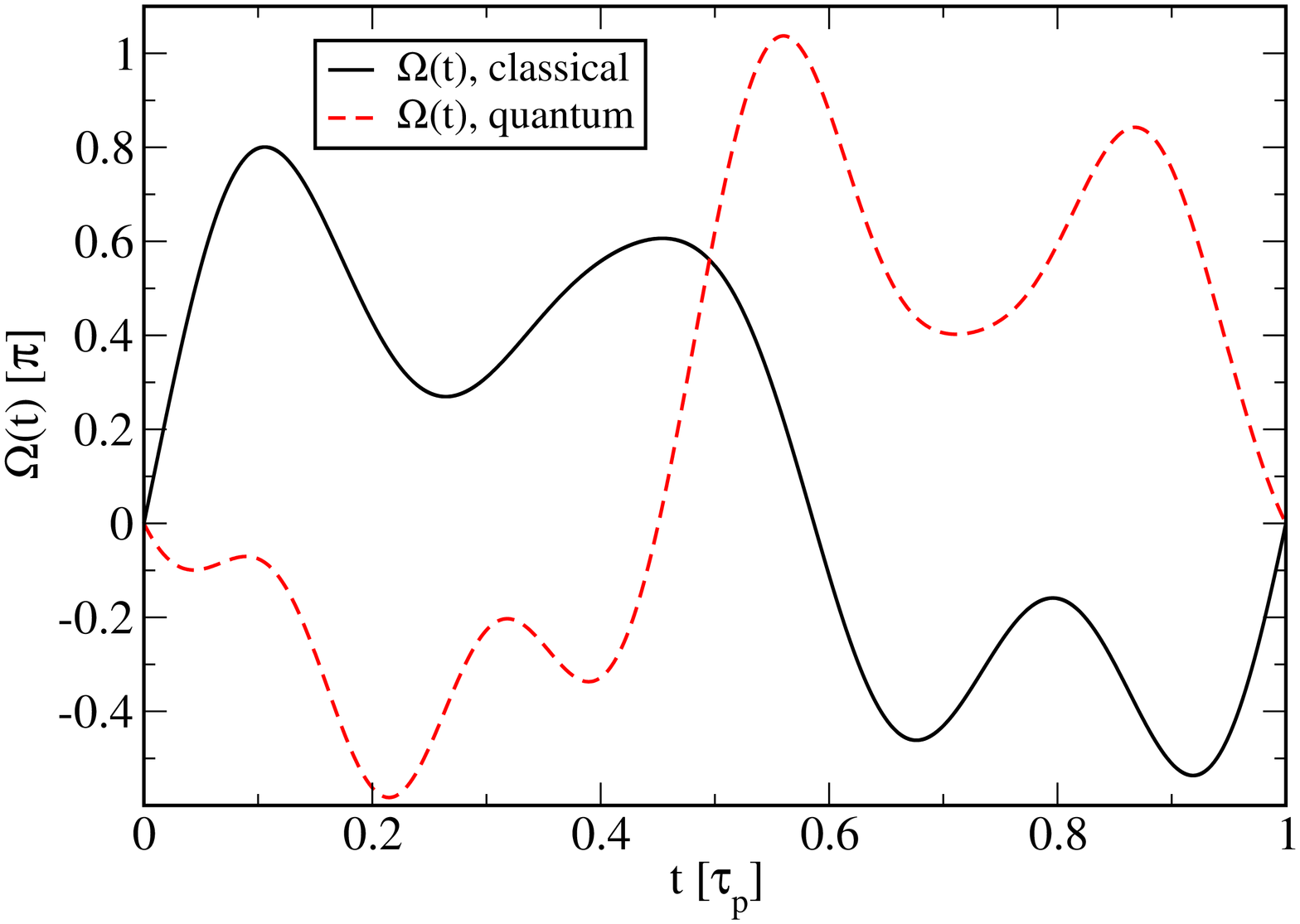}
  \caption{(Color online)  Minimized second order FM $\pi/2$ pulse suppressing pure
   dephasing with $\Omega(t)$ as in Eq.\ (\ref{eq:omega_fm}). The coefficients for 
   this pulse are given in Tab.\ \ref{tab:fm_2order_pi2}.
  \label{fig7}}
\end{figure}

\begin{figure}[htb]
 \includegraphics[angle=-90,width=\columnwidth,clip]{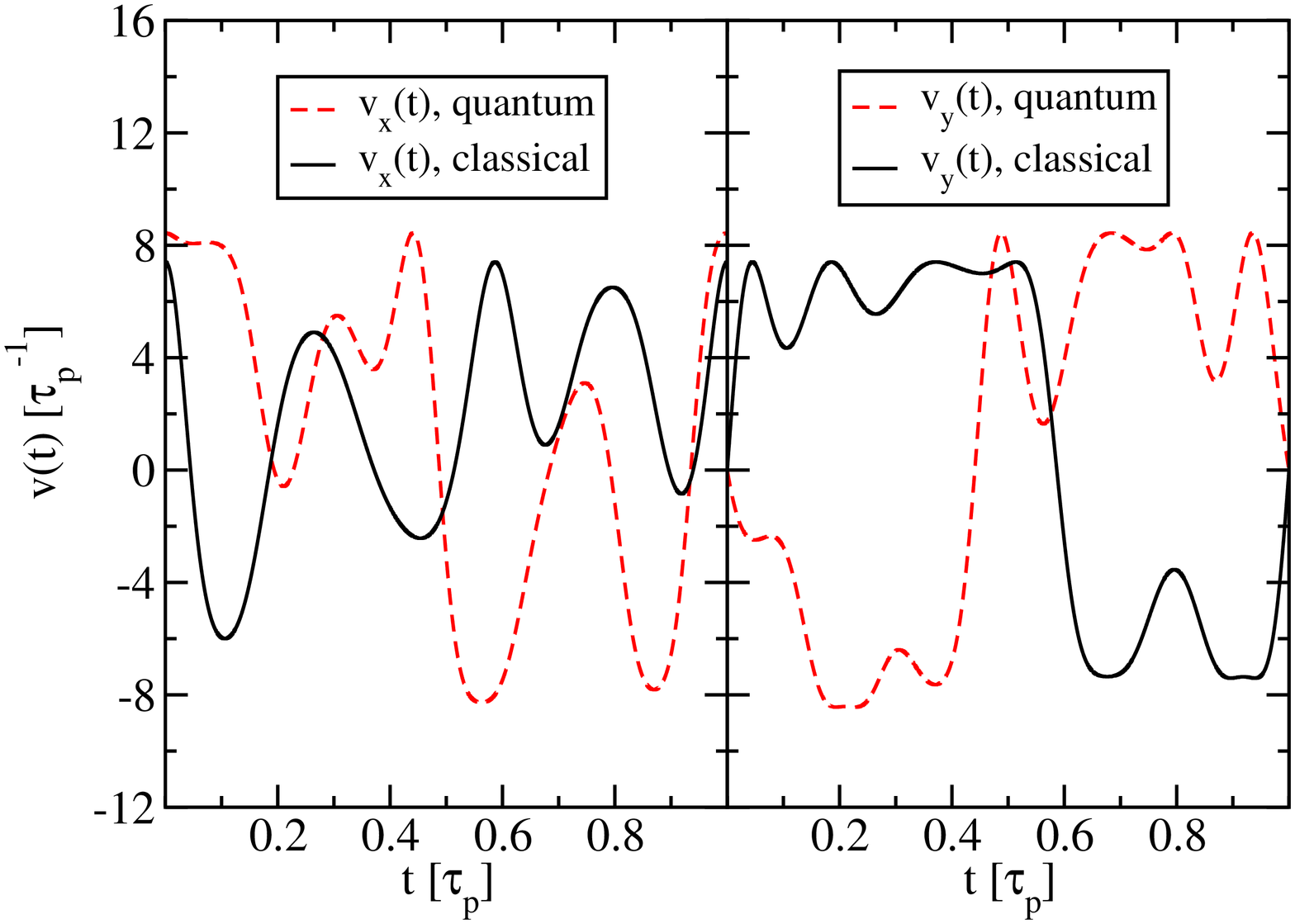}
  \caption{(Color online) Amplitudes for the same pulse as in Fig.\ \ref{fig7}.
  \label{fig8}}
\end{figure}

Solutions for second order pulses with frequency modulation are shown in Figs.\ \ref{fig5} and \ref{fig6} for $\pi$ pulses
and in Figs.\ \ref{fig7} and \ref{fig8} for $\pi/2$ pulse. In Figs.\ \ref{fig5} and \ref{fig7} the phases $\Omega(t)$
are shown while in Figs.\ \ref{fig6} and \ref{fig8} the corresponding amplitudes are plotted.
These pulses are minimized in the following way. Since it is important to realize pulses with 
small amplitudes for practical purposes, we studied sets of Fourier coefficients in \eqref{eq:omega_fm} 
with one coefficient more than necessary for the conditions \eqref{eq:1stFM} and \eqref{eq:2ndFM}.
We used this freedom to minimize the amplitude and carried this out for four different choices
of the Fourier coefficients. The solution with the minimal amplitude out of the four minima is shown.
The characteristics of the pulses are reported in Tabs.\ \ref{tab:fm_2order_pi} and \ref{tab:fm_2order_pi2}, respectively. We stress again that the classical case requires less coefficients than the quantum case. 
The maximal amplitudes of the classical pulses
are smaller than the amplitudes for the quantum pulse, as can be seen
comparing the values in the tables or the plots for $v_x$ and $v_y$. 
Moreover, the pulses suppressing classical noise have a simpler form.

Obviously, the solutions are symmetric because all sine coefficients in 
Tab.\ \ref{tab:fm_2order_pi} vanish to numerical
accuracy. This holds for the classical as well as for the quantum case. Deviations from this
symmetric shape in the results of Ref.\ \cite{fause12} are due to the lower precision of the
numerics used in this preceding article.

In addition to the tools employed already for AM pulses,
finding frequency modulated pulses requires to solve a system of three coupled ordinary differential
equations in each step of the multidimensional root search. This was done by using ``gsl\_odeiv2\_system'' with a
stepping of the type ``gsl\_odeiv2\_step\_rk4''. Thus a fourth order Runge-Kutta integration with adaptive step size
governed by the double step method to keep the local absolute error estimate in the
order of magnitude of $10^{-15}$.

\section{Amplitude and Frequency 	Modulated Pulses}
\label{sec:AM_FM_pulses}

Allowing for the modulation of amplitude and frequency leads
to a humongous parameter space. Thus we restrict ourselves to illustrating
that the frequency modulated pulses of the previous section can be
modified to account for smooth transients when the pulses are switched
on and off. Such transients are generic in experimental realizations.

\begin{figure}[htb]
 \includegraphics[width=\columnwidth,clip]{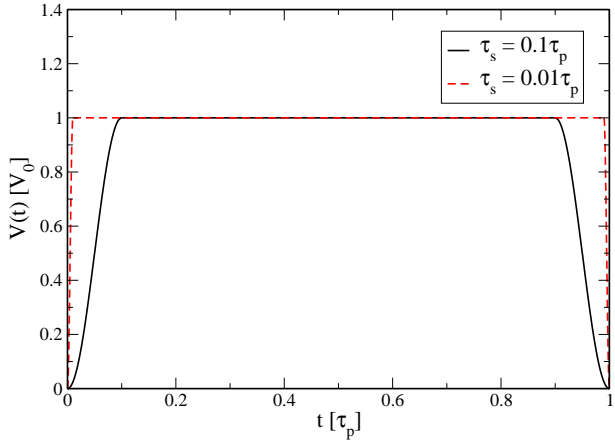}
  \caption{(Color online) A more realistic variation of the amplitude 
  $V(t)=V_0 f(t)$ according to
  	Eq.\ \eqref{eq:V0_AMFM} for frequency modulated pulses.
  Note the transients for switching the pulse on and off.}
  \label{fig:amfm-amp}
\end{figure}

To describe the transient region we define
\begin{align}
\label{eq:V0_AMFM}
f(t) := \left\{\begin{array}{rr} f_{\text{on}}(t)&\mbox{  $0\leq t<
      \tau_{\text{s}}$}\\ 
1 &\mbox{  $\tau_{\text{s}}\leq t<
  \tau_{\text{p}}-\tau_{\text{s}}$}\\
1-f_{\text{on}}[t-(1-\tau_{\text{s}})]&\mbox{
  $ \tau_{\text{p}}-\tau_{\text{s}}\leq t\leq \tau_{\text{p}} $} \end{array} \right.
\end{align}
where $f_{\text{on}}(t):= \sin^2\left(\pi t/2\tau_{\text{s}}\right)$.
The actual amplitude $V(t)$ is given by $V_0 f(t)$.
Then $\tau_{\text{s}}$ stands for the time needed to reach the amplitude $V_0$  from zero
amplitude or vice-versa, see Fig.\ \ref{fig:amfm-amp}.

In this section, we report the results for $\tau_{\text{s}}= 0.1
\tau_{\text{p}}$. In \ref{app:AM+FM}, results for other values of 
$\tau_{\text{s}}$ are included for comparison.
The parametrization of $\Omega(t)$ and ${\hat a}(t)$ are the same as in section \ref{sec:FM_pulses}.

\begin{figure}[htb]
 \includegraphics[angle=0,width=\columnwidth,clip]{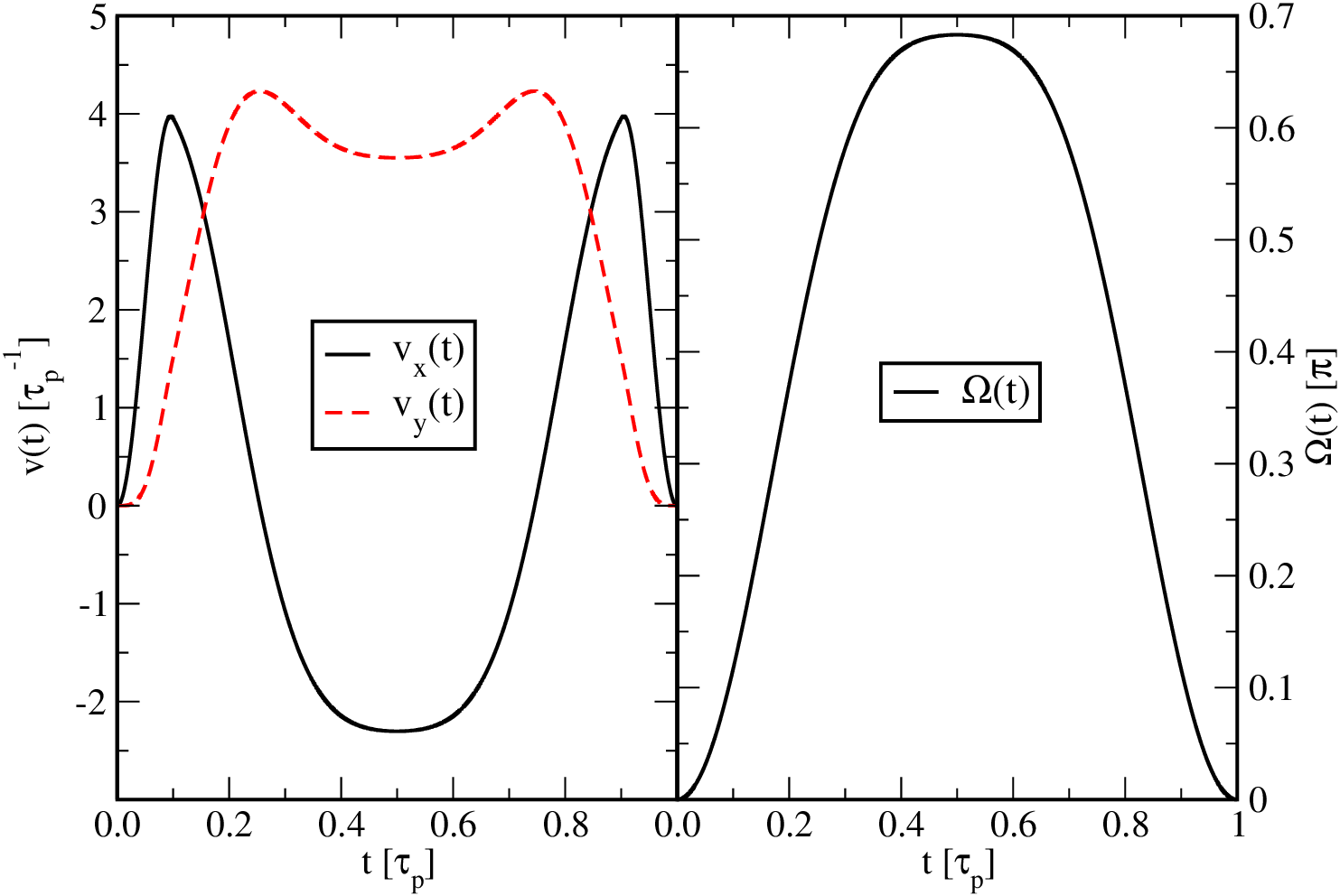}
  \caption{(Color online) AM+FM first order $\pi$ pulse suppressing pure dephasing
   for $\tau_{\text{s}}= 0.1 \tau_{\text{p}}$. The phase $\Omega(t)$ is parametrized as
    in Eq.\ (\ref{eq:omega_fm}). The coefficients of this pulse are given in Tab.\ \ref{tab:amfm_1order}.
  \label{fig9}}
\end{figure}

\begin{figure}[htb]
 \includegraphics[angle=0,width=\columnwidth,clip]{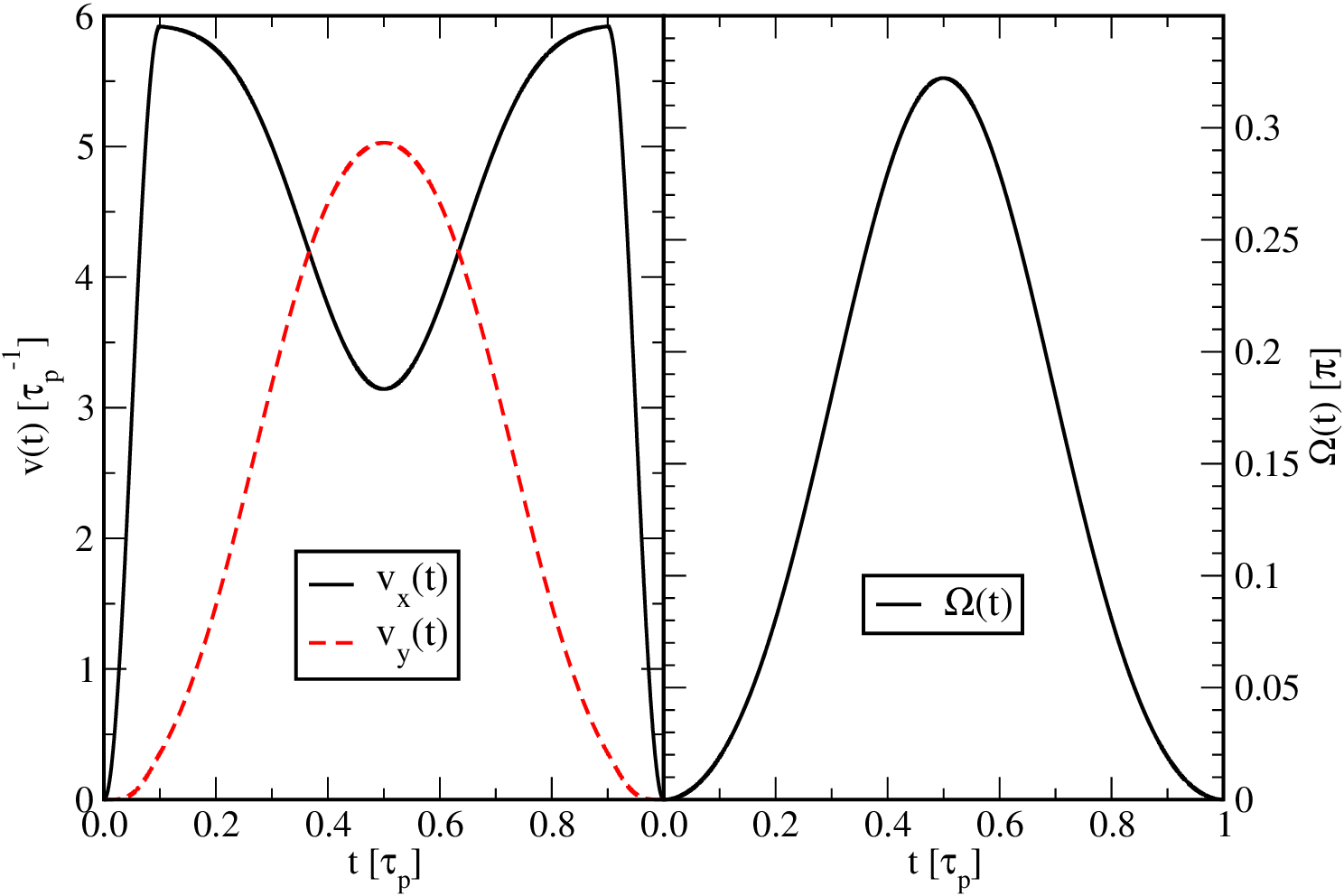}
  \caption{(Color online) AM+FM first order $\pi/2$ pulse suppressing pure dephasing
   for $\tau_{\text{s}}= 0.1 \tau_{\text{p}}$. The phase $\Omega(t)$ is parametrized as in Eq.\ (\ref{eq:omega_fm}). The coefficients of this pulse are given in Tab.\ \ref{tab:amfm_1order}.
  \label{fig10}}
\end{figure}

\begin{table}[htb]
 \centering
 \begin{tabular}{clcl}
\multicolumn{4}{c}{First order $\pi$ and $\pi/2$ AM+FM pulses}  \\
\multicolumn{2}{c}{$\pi$} &\multicolumn{2}{c}{$\pi/2$} \\
\hline								    
$V_0$  & 4.232216                               &  $V_0$  & 5.930552 \\
$b_1$  &  $0$   &  $b_1$  & $0$\\
$b_2$  &  -1.073059                            &  $b_2$  & -0.506131\\
$b_3$  &   $0$  &  $b_3$  & $0$\\
$b_4$  &  -0.233720                            &  $b_4$  &  0.053241\\
\end{tabular}
\caption{Parameters for the $\pi$ and $\pi/2$ AM+FM pulses 
suppressing pure dephasing with $\tau_{\text{s}}= 0.1
\tau_{\text{p}}$. The coefficients $b_n$ refer to Eq.\
  \eqref{eq:omega_fm}. The amplitudes $V_0$ are given in units of
  $1/\tau_\mathrm{p}$. The odd coefficients vanish numerically with in $10^{-11}$.
\label{tab:amfm_1order} }
\end{table}

Examples for first order AM+FM $\pi$ and $\pi/2$ pulses are
plotted in Figs.\ \ref{fig9} and \ref{fig10} and the coefficients of the parametrization can
be found in Tab.\ \ref{tab:amfm_1order}. The maximum amplitude of the AM+FM pulses is lower
than the amplitude of the continuous AM pulses and of the FM pulses derived for a quantum bath
\cite{pasin09a}. It is comparable to the amplitude of the well-known SCORPSE $\pi$ pulses
\cite{cummi00,cummi03} which is $7\pi/6\approx 3.7$ in units of $\tau_{\text{p}}^{-1}$.
In comparison to the pure FM pulses of first order shown in Figs.\ \ref{fig:1-PI-FM-DEKO}
and \ref{fig:1-PI2-FM-DEKO} with parameters given in Tab.\ \ref{tab:2-FM-DEKO} we clearly
see that a bit larger amplitudes are required for finite transients as was to be expected.

\begin{table}[htb]
 \centering
 \begin{tabular}{clcl}
\multicolumn{4}{c}{2nd order AM+FM $\pi$ and $\pi/2$ pulses}  \\
\multicolumn{2}{c}{$\pi$} &\multicolumn{2}{c}{$\pi/2$} \\
\hline								    
$V_0$  &  9.076304                             &  $V_0$  & 10.450781\\
$b_1$  &  $0$      &  $b_1$  & $0$\\
$b_2$  &  -0.436689                          &  $b_2$  &-0.123441\\
$b_3$  &   $0$  &  $b_3$  &$0$\\
$b_4$  &   0.305937                             &  $b_4$  &-0.130381\\
$b_5$  &  $0$   &  $b_5$  &$0$\\
$b_6$  &   -0.585209                         &  $b_6$  & -0.679511\\
$b_7$  &   $0$  &  $b_7$  & 	$0$\\
\end{tabular}
\caption{Parameters for the $\pi$ and $\pi/2$ AM+FM pulses suppressing pure dephasing by
 satisfying the first and second order  Eqs.\ \eqref{eq:1stFM} and \eqref{eq:2ndFM} for 
 a classical bath. The  coefficients $b_n$ refer to Eq.\  \eqref{eq:omega_fm}. The
  amplitudes $V_0$ are given in units of
  $1/\tau_\mathrm{p}$. The odd coefficients vanish numerically within $10^{-9}$.
\label{tab:amfm_2order} }
\end{table}

\begin{figure}[htb]
 \includegraphics[angle=0,width=\columnwidth,clip]{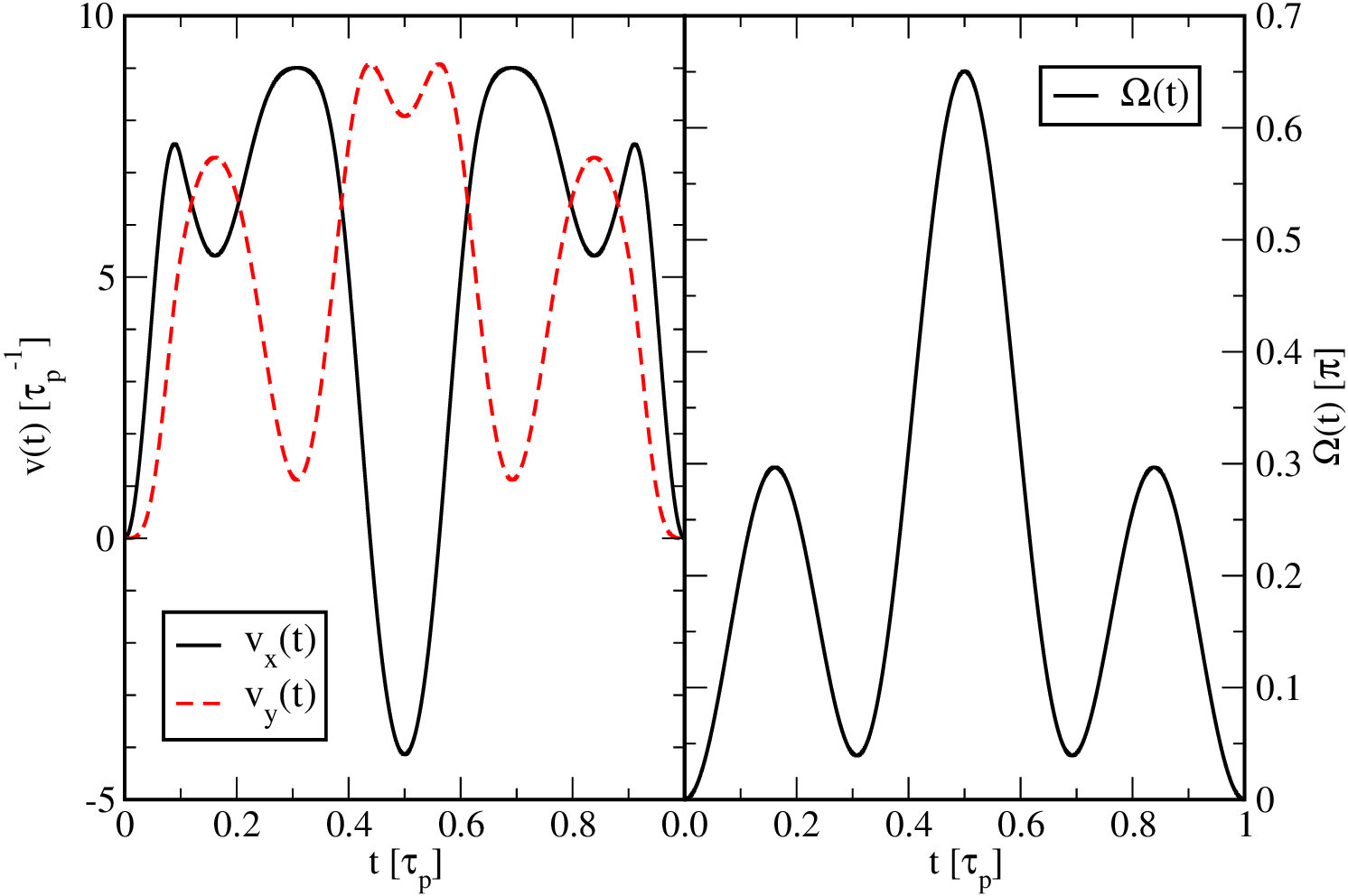}
  \caption{(Color online) Minimized second order AM+FM  $\pi$ pulse 
  suppressing pure dephasing with
    $\Omega(t)$ as in Eq.\ (\ref{eq:omega_fm}). The coefficients are given in Tab.\ \ref{tab:amfm_2order}.
  \label{fig11}}
\end{figure}

\begin{figure}[htb]
 \includegraphics[angle=0,width=\columnwidth,clip]{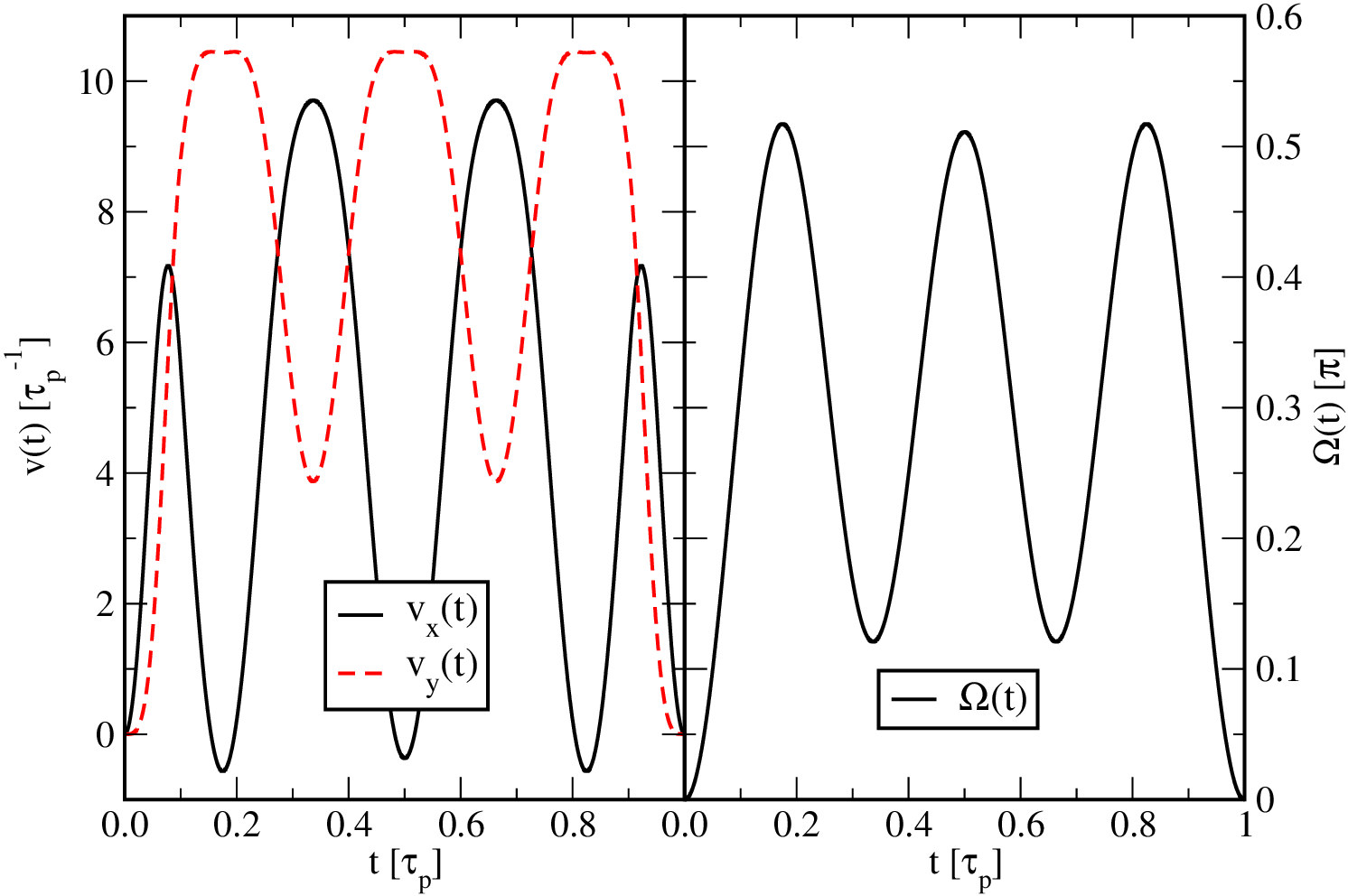}
  \caption{(Color online) Minimized second order AM+FM  $\pi/2$ pulse 
  suppressing pure dephasing with
    $\Omega(t)$ as in Eq.\ (\ref{eq:omega_fm}). The coefficients are given in Tab.\ \ref{tab:amfm_2order}.
  \label{fig12}}
\end{figure}

Analogous considerations hold for second order pulses. They are
plotted in Figs.\ \ref{fig11} and \ref{fig12}. 
Their coefficients are listed in Tab.\ \ref{tab:amfm_2order}.

\section{Frequency Modulated Pulses for General Decoherence}
\label{sec:gen_decoherence}

In section \ref{sec:FM_pulses} on frequency modulated pulses we pointed out that
one of their major assets is that they realize rotations about two independent spin
axes. Hence FM pulses may correct not only dephasing without spin flips
but also longitudinal relaxation including spin flips. We treat this
situation here. Thus we deal with the Hamiltonian in \eqref{eq:Hsystem}
where all three components $\eta_\alpha$ ($\alpha\in\{x,y,z\}$) of $\vec\eta$ are present.

We consider a bath with cylindrical symmetry generic for an NMR experiment where 
the $z$-axis of the spin is distinguished by a large magnetic field and the system is rotationally invariant around this axis. Considering rotations by $\pi$ and $\frac{\pi}{2}$ about $\sigma_z$ one easily finds
that the averages of $\vec\eta$ fulfill
\label{eq:expectval_deco}
\begin{align}
\overline{\eta}_x = \overline{\eta}_y &= 0
\end{align}
Additionally, second order pulse require the autocorrelation
\begin{equation}
\label{eq:autocorr_deco}
	g_{\alpha\alpha}\left(t=0\right) = \overline{\eta}_\alpha^2 +  s^2_\alpha + {\cal O}(t) .
\end{equation}
with $\alpha \in \{ x, y, z \}$. We
do not mention terms linear in $|t|$ because we aim at second order pulses at most.
In principle, the cross correlations $g_{\alpha\neq\beta}(t=0)$ may
also matter. But the cylindrical symmetry in combination with the antisymmetry of
$g_{xy}(-t)=-g_{xy}(t)$ imply that all cross correlations $g_{\alpha\neq\beta}(t=0)$ vanish.
In addition, $s_x^2=g_{xx}(0)=g_{yy}(0)$ is implied.

\begin{figure}[htb]
 \includegraphics[angle=0,width=\columnwidth,clip]{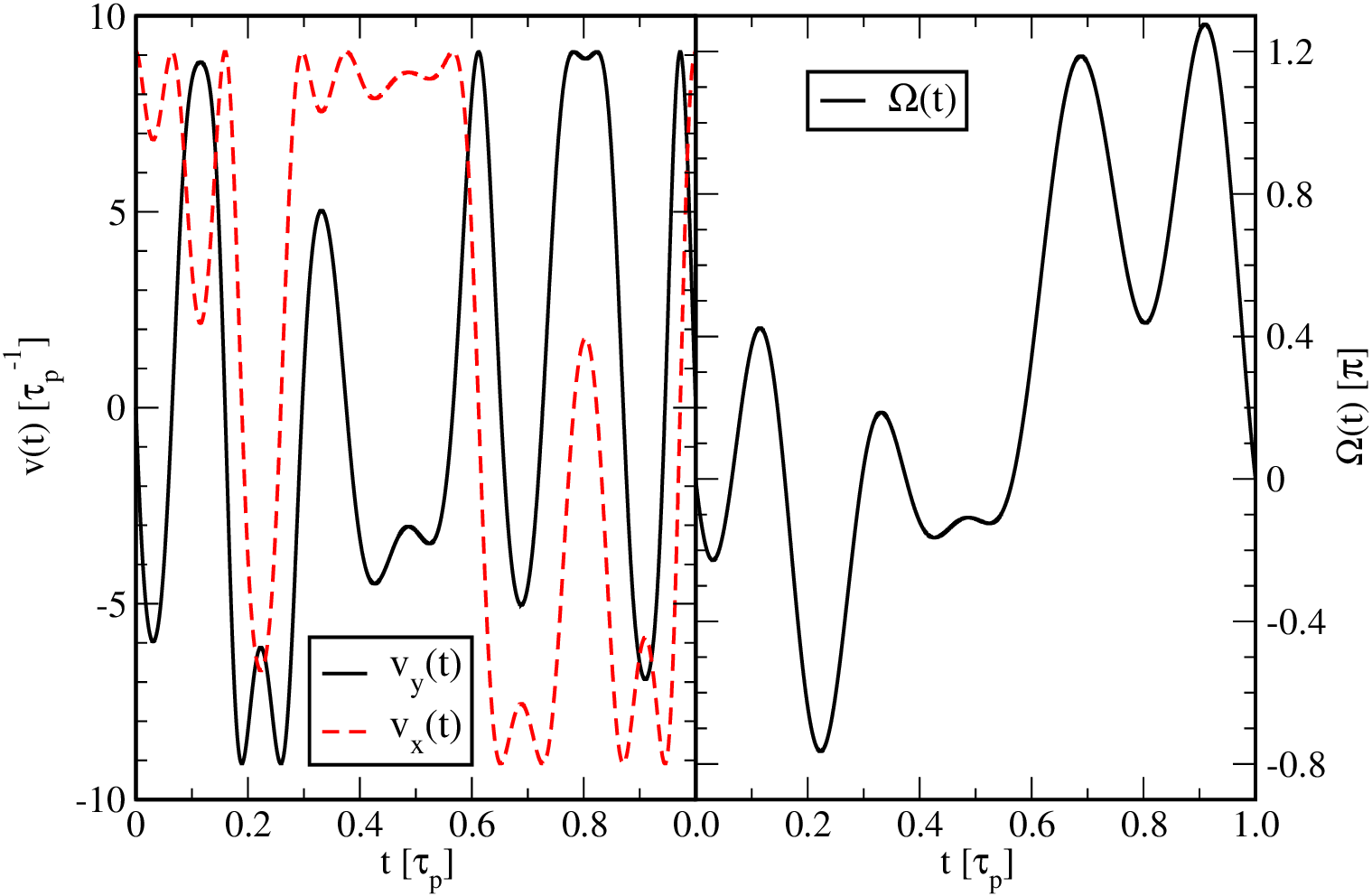}
  \caption{(Color online) Minimized second order FM $\pi$ pulse suppressing general decoherence
  with $\Omega(t)$ as in  Eq.\ (\ref{eq:omega_fm}). The coefficients  are given in Tab.\ \ref{tab:fm_2order_deco}.}
  \label{fig:2-PI-FM-DEKO}
\end{figure}

\begin{figure}[htb]
 \includegraphics[angle=0,width=\columnwidth,clip]{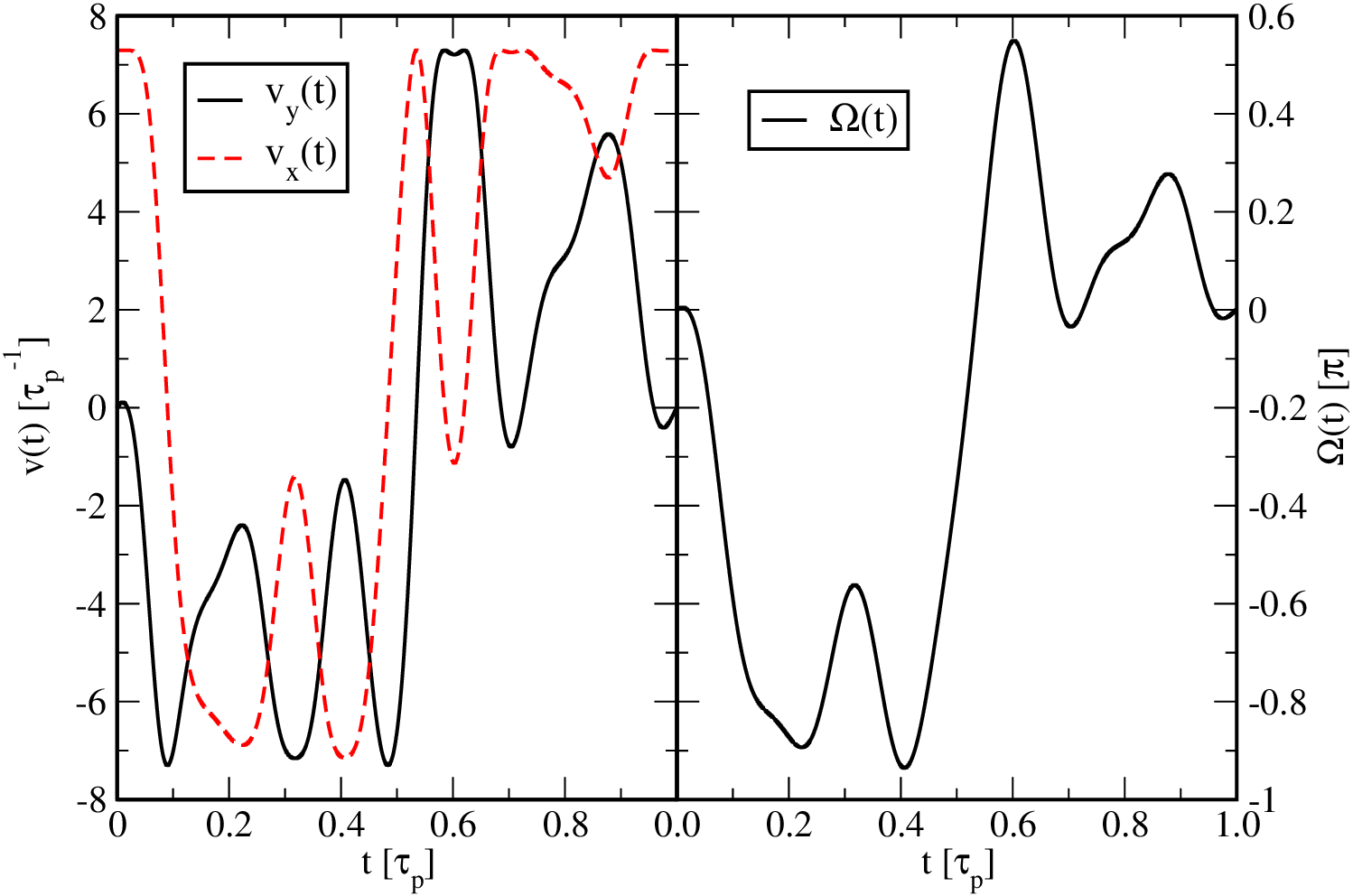}
  \caption{(Color online) Minimized second order FM $\pi/2$ pulse suppressing general decoherence
  with $\Omega(t)$ as in  Eq.\ (\ref{eq:omega_fm}). The coefficients  are given in Tab.\ \ref{tab:fm_2order_deco}.}
  \label{fig:2-PI2-FM-DEKO}
\end{figure}

On the basis of the
expectation values \eqref{eq:expectval_deco}, the vanishing of the first term,
see Eq.\ \eqref{eq:tildeH1},
requires the vanishing of the same integral equations as in the case of pure dephasing \eqref{eq:eta1_fm_deph} for frequency modulated pulses. Hence the same solutions result presented in the two previous sections.

In second order, we use the variances and vanishing cross correlations to conclude that 
the vanishing of 
\begin{equation}
\label{eq:2ndtildeH}
\tau_\text{p}\tilde H_2  = 
(\mu_{2,1}+\mu_{2,4})\sigma_x + (\mu_{2,2}+\mu_{2,5})\sigma_y + (\mu_{2,3}+\mu_{2,6})\sigma_z
\end{equation}
requires the coefficients $\mu_{2,i}$ to take the value zero
\begin{subequations}
\label{eq:eta2_deco_auto}
\begin{align}
\notag
\mu_{2,1} &= s^2_x \int\limits_0^{\tau_p}\int\limits_0^{t_1} [n_{yx,1}n_{zx,2} + n_{yy,1}n_{zy,2}\\ 
  &-\left( n_{zx,1}n_{yx,2} + n_{zy,1}n_{yy,2}\right) ]\mathrm{dt_2}\mathrm{dt_1}\\
  \notag
\mu_{2,2} &= s^2_x\int\limits_0^{\tau_p}\int\limits_0^{t_1} [n_{zx,1}n_{xx,2} + n_{zy,1}n_{xy,2}\\
  &-\left( n_{xx,1}n_{zx,2} + n_{xy,1}n_{zy,2}\right)] \mathrm{dt_2}\mathrm{dt_1}\\
  \notag
\mu_{2,3} &= s^2_x\int\limits_0^{\tau_p}\int\limits_0^{t_1} [n_{xx,1}n_{yx,2} + n_{xy,1}n_{yy,2}\\
  &-\left( n_{yx,1}n_{xx,2} + n_{yy,1}n_{xy,2}\right)] \mathrm{dt_2}\mathrm{dt_1}
 \end{align} 
 \begin{align} 
  \notag
  \mu_{2,4} &= \left(\overline{\eta}_z^2+s^2_z\right)\int\limits_0^{\tau_p}\int\limits_0^{t_1} [n_{yz}\left(t_1\right)n_{zz}\left(t_2\right)\\
  &\quad - n_{zz}\left(t_1\right)n_{yz}\left(t_2\right)] \mathrm{dt_2}\mathrm{dt_1}\\
  \notag
  \mu_{2,5} &= \left(\overline{\eta}_z^2+s^2_z\right)\int\limits_0^{\tau_p}\int\limits_0^{t_1} [n_{zz}\left(t_1\right)n_{xz}\left(t_2\right)\\
  &\quad - n_{xz}\left(t_1\right)n_{zz}\left(t_2\right)] \mathrm{dt_2}\mathrm{dt_1}\\
  \notag
  \mu_{2,6} &= \left(\overline{\eta}_z^2+s^2_z\right)\int\limits_0^{\tau_p}\int\limits_0^{t_1} [n_{xz}\left(t_1\right)n_{yz}\left(t_2\right)\\
  &\quad - n_{yz}\left(t_1\right)n_{xz}\left(t_2\right)] \mathrm{dt_2}\mathrm{dt_1}.
\end{align}
\end{subequations}
Note that we have to require that each $\mu_{2,i}$ vanishes even though they appear in pairs
in front of the Pauli matrices in \eqref{eq:2ndtildeH} because we do not want to make assumptions
of the relative size of $s_x^2$ and $\overline{\eta}_z^2+s^2_z$.

The parametrization of $\Omega(t)$ and ${\hat a}(t)$ are chosen as before in sections \ref{sec:FM_pulses}
and \ref{sec:AM_FM_pulses}. 
Solutions to the conditions $\mu_{1,i}=0$ and $\mu_{2,i}=0$ are shown in Fig.\ \ref{fig:2-PI-FM-DEKO} for a $\pi$ pulse and in Fig.\ \ref{fig:2-PI2-FM-DEKO} for a $\pi/2$ pulse.  The parameters of these pulses are reported in Tab.\ \ref{tab:fm_2order_deco}. The amplitudes of these pulses have been minimized as described before.

Comparing the amplitudes in Tab.\ \ref{tab:fm_2order_deco} with those for classical pulses
in Tabs.\ \ref{tab:fm_2order_pi} and \ref{tab:fm_2order_pi2}, it is clear that the suppression of
general decoherence requires higher amplitudes or longer pulses, respectively. But the increase in amplitude is not very large. For the $\pi$ pulse,
the amplitude is increased by 12\% and for the $\pi/2$ pulse 
it is even lowered by 1\%. This finding appears 
contradictory at first sight because more conditions have to 
be fulfilled for general decoherence. The contradiction
is resolved by the observation that we consider more coefficients
in the construction of the pulse suppressing general decoherence than
we do in the construction of the pulse suppressing pure dephasing,
see Tabs.\ \ref{tab:fm_2order_pi2} and \ref{tab:fm_2order_deco}.
Of course, the $\pi/2$ pulse suppressing general decoherence fulfills
also the conditions for the suppression of pure dephasing.
It is remarkable that the additional conditions for general decoherence
can be fulfilled at only moderate additional effort. Hence one may realize
a pulse which not only suppresses transversal decoherence but also  longitudinal decoherence.

Another intriguing possibility is to use the pulses found as replacement for
XY4 and XY8 cycles \cite{mauds86,gulli90}. The main difference between the
net effect of XY4 to the $\pi$ pulse suppressing general decoherence in
Fig.\ \ref{fig:2-PI-FM-DEKO}  and Tab.\ \ref{tab:fm_2order_deco} is that XY4 is
designed to be a no-operation (NOOP) sequence while the $\pi$ pulse realizes a spin flip.
But if the $\pi$ pulses is applied a second time in time-reversed order $\Omega(\tau)\to\Omega(\tau_\text{p}-\tau)$
it realizes again a second order $\pi$ pulses. Thus the effect of both pulses back-to-back
is a $2\pi$ pulse which reduces to a mere phase factor $-1$ without effect on the actual 
spin state. Hence, such a composite pulse of length $2\tau_\text{p}$
\begin{equation}
\pi\Big|_\text{FM,forward} - \pi\Big|_\text{FM,backward}
\end{equation}
is symmetric and corresponds to an XY8 cycle in the sense that it suppresses
spin dephasing and spin flips; thus it suppresses general decoherence up to
second order. The intriguing aspect is that this composite pulse is an always-on pulse
and does not consist of 8 individual pulses. Thus its amplitude is much lower than
the amplitude needed in an XY8 cycle. For given pulse length one can reduce the
total energy $\propto V_0^2\tau_\text{p}$ needed for the coherent control.
 For given maximal amplitude the cycle can be performed much more rapidly.
 This opens a promising route to more effective control calling for experimental
 verification.

\begin{table}[htb]
 \centering
 \begin{tabular}{clcl}
\multicolumn{4}{c}{\small{2nd order FM $\pi$ and $\pi/2$ pulses for general decoherence}}\\
\multicolumn{2}{c}{$\pi$} & \multicolumn{2}{c}{$\pi/2$}\\
\hline
$V_0$ &  $9.079728$ & $V_0$ & $7.297361$\\ 
$b_1$ & $1.818085$ &  $b_1$ & $-1.793195$\\ 
$b_2$ & $0.514273$ & $b_2$ & $0.223583$\\ 
$b_3$ & $-0.231238$ & $b_3$ & $0.221590$\\ 
$b_4$ & $-0.220323$ & $b_4$ & $0.324311$\\ 
$b_5$ &  $0.014857$ & $b_5$ & $-0.579783$\\ 
$b_6$ &  $0.508720$ & $b_6$ & $0.272144$\\ 
$b_7$ & $-0.439837$ & $b_7$ & $0.507358$\\ 
$b_8$ & $-0.816150$ & $b_8$ & $-0.119786$\\ 
$b_9$ & $-0.332547$ & $b_9$ & $-0.011429$\\ 
$b_{10}$& $-0.846412$ &  $b_{10}$ & $0.069581$\\
$b_{11}$ & $-0.249481$ & $b_{13}$ & $0.219071$
\end{tabular}
\caption{
Parameters for the minimized FM $\pi$ and $\pi/2$ pulses suppressing general decoherence. The 
  coefficients $b_n$ refer to Eq.\ \eqref{eq:omega_fm}. 
  The amplitudes $V_0$ are given in units of $1/\tau_\mathrm{p}$.}
\label{tab:fm_2order_deco}
\end{table}

\section{Conclusions}
\label{sec:conclusions}

Coherent control of quantum systems is a very active field of current research.
The simplest quantum system to be controlled generally is a two-level system
which can be seen as a spin $S=1/2$. In particular, the coherent control
of such spins is at the very heart of magnetic resonance techniques, where
nuclear spins are manipulated, and of quantum information processing, where
quantum bits are manipulated.
Previously, pulses were designed to  suppress the influence of a noisy quantum environment.
Although this is the most general case it is cumbersome because many conditions
have to be  fulfilled and thus the necessary pulses are fairly complicated.
They need relatively large amplitudes or they are relatively long.
Both properties limit their practical relevance.

In practice, however, very often the environmental noise is dominated by
classical fluctuations, for instance because it results from macroscopic
degrees of freedom at relatively high temperature (room temperature).
This led us to study pulses which suppress classical noise. 
Such pulses are subject to less conditions so that they can be
simpler and of lower amplitudes or their duration $\tau_\text{p}$ can be shorter.
We studied such pulses analytically by a systematic expansion in $\tau_\text{p}$.
Pulses with amplitude modulation, which rotate the spin about a single
fixed axis, and pulses with frequency modulation, which rotate the spin about
a continuously varying axis in the $xy$-plane of the spin directions, are considered.

A purely dephasing bath is relevant if the energy splitting between the two
levels is so large that the rotating frame approximation is applicable.
Experimentally, the dephasing dominates if $T_2$ is  much smaller than $T_1$.
For this situation we presented first and second order pulses
based on amplitude or on frequency modulation, respectively.
We explicitly presented pulse shapes for $\pi$ and $\pi/2$ pulses which
either flip the spin between up and down or which rotate them by
90$^\circ$ from the $z$-axis.
In all pulses suppressing classical noise we confirmed that their
amplitudes are lower than for their previously known quantum counterparts.
The only exception are first order pulses. In this order the quantum 
character is not manifest in the conditions so that they are identical
for classical and for quantum baths.

Furthermore, we showed that combinations of amplitude and frequency
modulation can also be treated. In particular, finite transients
in the switching processes of amplitudes can be accounted for.
Thus imperfections in the switching can be considered and they do not
pose a conceptual problem.

Intriguingly, we could furthermore establish the existence of pulses
which suppress classical general decoherence. This means that not only
transversal dephasing but also longitudinal spin relaxation relying
on spin flips can be suppressed. Amplitude modulation is not sufficient
to this end because one needs at least two independent axes of rotation to
suppress all kinds of spin errors. But frequency modulated pulses can do the job.

We found that the necessary amplitudes are at worst only moderately larger 
than for the suppression of pure dephasing alone. 
Thereby, we propose a single shaped pulse which has similar properties
as cycles of pulses. In particular, a composite pulse of two $\pi$ 
pulses, which suppress general decoherence, can replace the
well-known XY8 cycle. The asset of the composite pulse we are advocating here
is that it is an always-on pulse. Hence the required amplitudes are 
much lower than those required for a cycle of very short pulses.

Of course, further research is called for. On the experimental side, studies
of the performance of the proposed pulses are called for. A key question
is whether the proposed shapes can be realized reliably enough to 
reach the predicted suppression of the noisy baths.

Theoretically, the question of the robustness of the proposed pulses
towards imperfections in their realizations deserves investigations.
For instance, simulations of the pulses in various baths are called for
to guide experiment. In parallel, we believe that the necessary amplitudes can 
still be reduced by further minimization.

Finally, we point out that analogous expansions for coupled
two-level systems have hardly attracted attention so far in spite
of their relevance of two-qubit gates in quantum information processing.

\section{Acknowledgements}
We acknowledge financial support of the DFG in Project UH 90/5-1.

\appendix

\section{Pulses with Amplitude and Frequency Modulation}
\label{app:AM+FM}

In section \ref{sec:AM_FM_pulses}, we showed an example of
two FM pulses with angle $\pi$ and $\pi/2$ and finite transients
of amplitude for switching on and off. Each transient takes
a fraction $\tau_{\text{s}}/\tau_\text{p}$ of the total time $\tau_{\text{p}}$ of the
pulse. Here we provide further $\pi$ pulses with shorter transient time
in Tab.\  \ref{tab:amfm_1order_taus} in first order and 
in Tab.\  \ref{tab:amfm_2order_taus} in second order.
The parameters for the first order pulses are to be compared to those
reported in  Tab.\ \ref{tab:amfm_1order}. 
Remarkably, the amplitude does not increase monotonically on 
increasing $\tau_\text{s}/\tau_{\text{p}}$ in Tab.\ \ref{tab:amfm_1order_taus}.

The second order parameters
illustrate that amplitude and frequency modulation can also be combined
in second order. The parameters can be compared to those of pure FM pulses
given in Tab.\ \ref{tab:fm_2order_pi}. 

\begin{table}
\centering
 \begin{tabular}{clcl}
\multicolumn{4}{c}{1st order  AM+FM  $\pi$pulses}  \\
\multicolumn{2}{c}{$\tau_{\text{s}}=0.01 \tau_{\text{p}}$} &\multicolumn{2}{c}{$\tau_{\text{s}}=0.001 \tau_{\text{p}}$} \\
\hline								    
$V_0$	&	$3.907279$                          & $V_0$	&	$4.016102$\\
$b_1$	&	$0$ & $b_1$	&	$0$\\
$b_2$	&	$-0.892324$                      & $b_2$	&	$-0.791296$\\
$b_3$	&	$0$   & $b_3$	&	$0$\\
$b_4$	&	$-0.196455$                      & $b_4$	&	$-0.040110$\\
\end{tabular}
\caption{Parameters of AM+FM first order $\pi$ suppressing pure dephasing for two switching times $\tau_{\text{s}}$. The dimensionless
  coefficients $b_n$ refer to Eq.\
  \eqref{eq:omega_fm}. The amplitudes $V_0$ are given in units of
  $1/\tau_\mathrm{p}$. The odd coefficients vanish within $10^{-11}$.
\label{tab:amfm_1order_taus} }
\end{table}

\begin{table}
\centering
 \begin{tabular}{clcl}
\multicolumn{4}{c}{2nd order  AM+FM  $\pi$pulses}  \\
\multicolumn{2}{c}{$\tau_{\text{s}}=0.1 \tau_{\text{p}}$} &\multicolumn{2}{c}{$\tau_{\text{s}}=0.01 \tau_{\text{p}}$} \\
\hline								    
$V_0$    &    $9.076304$ & $V_0$    &    $8.486171$\\
$b_2$    &    $-0.436689$ & $b_2$    &    $-0.309163$\\
$b_4$    &    $0.305937$ & $b_4$    &    $0.507966$\\
$b_6$    &    $-0.585209$ & $b_6$    &    $-0.437161$\\
\end{tabular}
\caption{Parameters of AM+FM second order $\pi$ suppressing pure dephasing for two switching times $\tau_{\text{s}}$. The dimensionless
  coefficients $b_n$ refer to Eq.\
  \eqref{eq:omega_fm}. The amplitudes $V_0$ are given in units of
  $1/\tau_\mathrm{p}$. The odd coefficients vanish within $10^{-9}$.
\label{tab:amfm_2order_taus} }
\end{table}

\section{Rotation Matrix}
\label{app:rotmatrix}

For the derivation of the matrix $D_{\hat{a}}(-\psi)$ we refer the reader to Ref.\ \cite{pasin09a}.  We obtain the matrix \eqref{eq:matrix} below, where the time dependencies of $\psi(t)$ and $\hat{a}(t)$ are omitted for clarity.
The matrix elements $[D_{\hat{a}}(-\psi)]_{\alpha\beta}$ define the quantities $n_{\alpha\beta}$
where we identify $x$ with $1$, $y$ with $2$, and $z$ with $3$.
\begin{onecolumn}
\begin{align}
D_{\hat{a}}(-\psi) = \left(\begin{array}{ccc}
\cos \psi + (1- \cos \psi)a^2_{x} 	   & a_{z} \sin \psi + (1 - \cos \psi) a_{x} a_{y}  & -a_{y} \sin \psi + (1 - \cos \psi) a_{x} a_{z} \\
-a_{z} \sin \psi + (1- \cos \psi)a_{x} a_{y} &     \cos \psi + (1 - \cos \psi) a^2_{y}    & a_{x} \sin \psi + (1 - \cos \psi) a_{y} a_{z} \\
a_{y} \sin \psi + (1- \cos \psi)a_{x} a_{z}  & -a_{x} \sin \psi + (1 - \cos \psi) a_{y} a_{z} & \cos \psi + (1 - \cos \psi) a^2_{z} 
                      \end{array}\right) 
\label{eq:matrix}
\end{align}
\end{onecolumn}

\begin{twocolumn}

\end{twocolumn}

\end{document}